\documentclass{article}

\usepackage{arxiv}

\usepackage[utf8]{inputenc} 
\usepackage[T1]{fontenc}    
\usepackage{url}            
\usepackage{booktabs}       
\usepackage{amsfonts}       
\usepackage{nicefrac}       
\usepackage{microtype}      
\usepackage{lipsum}		
\usepackage{graphicx}
\usepackage{natbib}
\usepackage{doi}
\usepackage{newtxtext}
\usepackage{newtxmath}
\usepackage{authblk}
\usepackage{tabu}
\usepackage{multirow}

\title{Phase-based analysis and control of supersonic turbulent cavity flows}

\author[1]{Vedasri Godavarthi$^1$\thanks{Corresponding author: vedasrig@g.ucla.edu}}
\author[2]{Yoji Kawamura}
\author[3]{Lawrence S. Ukeiley}
\author[4]{Louis N. Cattafesta III}
\author[1]{Kunihiko Taira}
\affil[1]{Department of Mechanical and Aerospace Engineering, University of California, Los Angeles, CA 90095, USA}
\affil[2]{Center for Mathematical Science and Advanced Technology, Japan Agency for Marine-Earth Science and Technology, Yokohama 236-0001, Japan}
\affil[3]{Department of Mechanical and Aerospace Engineering, University of Florida, Gainesville, FL 32611, USA}
\affil[4]{Department of Mechanical, Materials, and Aerospace Engineering, Illinois Institute of Technology, Chicago, IL, 60616, USA}

\hypersetup{
pdftitle={A template for the arxiv style},
pdfsubject={q-bio.NC, q-bio.QM},
pdfauthor={David S.~Hippocampus, Elias D.~Striatum},
pdfkeywords={First keyword, Second keyword, More},
}

\begin{document}
\maketitle

\begin{abstract}
	We present a phase-based framework for reducing the pressure fluctuations within a spanwise-periodic supersonic turbulent cavity flow with an incoming free-stream Mach number of 1.4 and a depth-based Reynolds number of $10^4$. Open cavity flows exhibit large fluctuations due to the feedback between the shear layer instabilities and the acoustic field. The dominant flow physics includes the formation, convection, and impingement of large-scale spanwise-oriented vortical structures. We formulate a flow control strategy to effectively modify the vortex convection frequency, thereby disrupting the feedback loop and suppressing pressure fluctuations within the cavity. We implement a phase-reduction approach to identify the flow response about the time-varying convective process by defining a phase variable using dynamic mode decomposition. Three-dimensional impulse perturbations are introduced from the cavity leading edge to characterize phase response in terms of advancement or delay of convection. We perform open-loop flow control through unsteady blowing using actuation frequencies slightly different from the vortex convection frequency to disrupt the feedback loop. After designing a phase-sensitivity-based actuation waveform optimized for quick flow modification, we investigate the speed of fluctuation reduction and compare it to a sinusoidal waveform. At a spanwise actuation wavenumber of $\beta=2\pi$, both perform similarly, achieving a 46\% reduction of pressure fluctuations within five convective times. At $\beta=\pi$, the phase-sensitivity-based actuation performs better by achieving a 40\% reduction compared to the 30\% with a sinusoidal waveform within six convective times. This study shows the potential of phase-based analysis for timing-based flow control of unsteady turbulent flows.
\end{abstract}


	\section{Introduction}
	\label{sec:intro}
	
Flow control of shear layer instabilities in open cavity flows has been of interest for several decades as they are encountered in several engineering applications, such as landing gear wells, bays of aircraft, and combustors in automobiles. High-speed cavity flows exhibit self-sustaining fluctuations driven by a feedback loop between acoustic waves and shear layer instabilities, compounded by broadband turbulence. These fluctuations can cause structural fatigue, increase drag, and are a source of noise production \citep{mcgregor1970drag, dix2000experimental}. Hence, designing effective control to minimize the pressure fluctuations based on the oscillatory flow states is critical. 

The shear layer oscillations in the cavity flow arise due to the flow-acoustic feedback in the cavity. The shear layer emanating from the leading edge of the cavity rolls up into large spanwise vortical structures, which convect and impinge on the aft wall of the cavity, generating strong acoustic waves. These acoustic waves then travel upstream, affecting the receptive shear layer at the leading edge, completing the natural feedback loop. The resonant cavity tones arising from this feedback loop are characterized by the Rossiter frequencies \citep{heller1971flow}. Numerous passive and active control strategies have been aimed at suppressing pressure fluctuations by modifying the shear layer flow features \citep{colonius2001overview,rowley2006dynamics,cattafesta2008active}. A few instances of passive control strategies include lifting the shear layer at the leading edge \citep{sarno1994suppression} and placing a cylinder at the leading edge to reduce fluctuations in the cavity \citep{mcgrath1996active,ukeiley2004suppression}. 

Active control of high-speed cavity flows has been of significant interest in promoting the adaptability of control to a wide range of free-stream flow features \citep{cattafesta2008active}. Parametric studies using open-loop forcing introduced steady blowing at the leading edge and showed the efficacy of three-dimensional steady injection in suppressing pressure fluctuations in cavity flows compared to the two-dimensional injection \citep{lusk2012leading,george2015control,zhang2019suppression,bacci2023hilbert}. Further compared to steady blowing, unsteady blowing is more effective at suppressing pressure fluctuations and has been widely used for cavity flow control \citep{rowley2006dynamics, cattafesta2008active,zhuang2006supersonic,ali2010control,webb2017control}. Unsteady actuation at detuned frequencies relative to the dominant Rossiter tones is considered herein to suppress the cavity tones, resulting in a reduction of pressure fluctuations.

The actuation frequency is an important parameter in these works and motivated several studies. Actuation at an order of magnitude lower frequency than the Rossiter modes is not effective in suppressing the cavity tones \citep{sarno1994suppression}. The unsteady actuation at a frequency of the order of the dominant tones is shown to suppress some tones at a sufficient actuation amplitude, while in some scenarios, also results in lock-on to other resonant tones \citep{cattafesta1997active,samimy2004exploring,takahashi2011progress}. \cite{williams2007supersonic} performed open-loop flow control of the supersonic cavity using pulsed-blowing and varied the actuation frequency and magnitude. They observed the suppression of the cavity tone when the actuation frequency is in between the Rossiter modes and an enhancement when the actuation frequency is close to a Rossiter mode due to lock-on resonance. On the other hand, high-frequency actuation at an order of magnitude higher than Rossiter modes is effective for suppression \citep{stanek2001suppression,sipp2012open,kreth2020using}. However, such high-frequency actuation for high-speed flows is a challenge for current available actuators \citep{cattafesta2011actuators}. In addition to open-loop control strategies, there also have been studies that performed closed-loop control based on reduced order modeling of cavity flows. Feedback controllers based on linear models and reduced order models of cavity flow have also been demonstrated to significantly suppress the dominant cavity tones~\citep{rowley2006linear,yan2006experimental,samimy2007feedback,illingworth2012feedback}. Closed-loop control studies have been very promising for cavity flow control, while their application to supersonic high-speed cavity flows requires further additional progress.

More recently, resolvent analysis has been used to identify the flow response to harmonic forcing for high-speed open cavity flows \citep{sun2020resolvent,liu2021unsteady,godavarthi2024windowed}. \cite{liu2021unsteady} performed resolvent-based open-loop flow control for supersonic cavity flow by identifying a spanwise actuation wavenumber and frequency that results in the most amplified response in the shear layer and obtained a significant 52\% suppression of pressure fluctuations on the cavity walls. The resolvent-based control studies characterize the system response to the harmonic perturbations about the time-averaged base-state. Given the highly oscillatory nature of cavity flows, designing a control strategy that accounts for these unsteady base states remains an open question. This work aims to design a flow-control strategy by characterizing the perturbation dynamics of the oscillatory base states of supersonic turbulent cavity flow. To this end, we leverage phase reduction analysis. 

The phase reduction approach characterizes the perturbation evolution about a periodically varying base state and has been widely used in several biological and chemical systems \citep{kuramoto1984chemical,strogatz1994chaos,nakao2016phase}. This approach has recently seen increased application in fluid systems, predominantly periodic fluid flows \citep{kawamura2013collective,kawamura2015phase,taira2018phase,skene2022phase,loe2023controlling}. Phase reduction analysis transforms the high-dimensional perturbation dynamics about a periodically varying base state to a single scalar phase variable. It has been used to characterize the lock-in conditions and control the periodic vortex shedding over a cylinder \citep{taira2018phase, khodkar2020phase,khodkar2021phase}, airfoil \citep{nair2021phase,kawamura2022adjoint} and flat plate \citep{iima2024optimal}. Phase-based analysis has been used to design an optimal actuation waveform that can quickly lock-in to a forcing frequency slightly different than the dominant frequency for periodic flows. This has been demonstrated for rapid modification of flow physics for laminar periodic flows \citep{godavarthi2023optimal,loe2023controlling}, including periodic airfoil wakes and transient gust-airfoil interactions \citep{fukami2024data}.

While phase reduction analysis is limited to periodic flows, a few studies have considered phase reduction for simpler chaotic nonlinear oscillators \citep{josic2001phase,schwabedal2012optimal,tonjes2022phase}. Recently, \cite{kim2024influence} extended phase reduction analysis for almost periodic fluid flows using an ensemble-averaged method. Motivated by the recent applications of phase reduction analysis to achieve frequency modification and using the insights from resolvent analysis-based control \citep{liu2021unsteady}, we aim to suppress the dominant cavity tone by actuating at a slightly different frequency using phase-based control. We first identify the phase description of supersonic turbulent cavity flow to capture the dominant process of spanwise vortex convection in the shear layer. Further, we extend the phase reduction approach to turbulent fluid flows by identifying the average response to perturbations in terms of delay/advancement of the vortex convection. We then perform open-loop forcing at an actuation frequency slightly different from the dominant vortex convection frequency to suppress the dominant cavity tone. Since the feedback mechanism is responsible for the cavity flow oscillations, we hypothesize that modification of the time scale of vortex convection results in disruption of this feedback loop, thereby suppressing the pressure fluctuations in the cavity. Further, we also design a phase-sensitivity-based actuation waveform to quickly modify the frequency of vortex convection in the cavity and compare the control performance and fluctuation reduction speed with a sinusoidal waveform. This work demonstrates the applicability of timing-based flow control for oscillatory turbulent flows. 

The outline of the paper is as follows. The numerical setup is discussed in section.~\ref{sec:setup}. Section~\ref{sec:methodology} describes the flow physics of uncontrolled cavity flows and the phase description. The methodology of phase reduction and phase-sensitivity-based waveform is provided in section~\ref{sec:phase_reduction}. The results are discussed in section~\ref{sec:results} and the conclusions in section~\ref{sec:conclusions}.


\section{Problem setup}
\label{sec:setup}
We aim to reduce pressure fluctuations in supersonic turbulent flow over an open rectangular cavity. We examine the spanwise-periodic flow over a rectangular cavity of a length-to-depth ratio of $L/D=6$, with incoming free-stream Mach number, $M_\infty=U_\infty/a_\infty=1.4$ and a Reynolds number, $Re=U_\infty D/\nu_\infty= 10000$, where $U_\infty,\,a_\infty,\,\nu_\infty$ are the free-stream velocity, speed of sound and kinematic viscosity, respectively. The depth of the cavity $D$ is chosen as the characteristic length scale for all the spatial dimensions.

 \begin{figure}
\centering
    \includegraphics[width=0.8\textwidth]{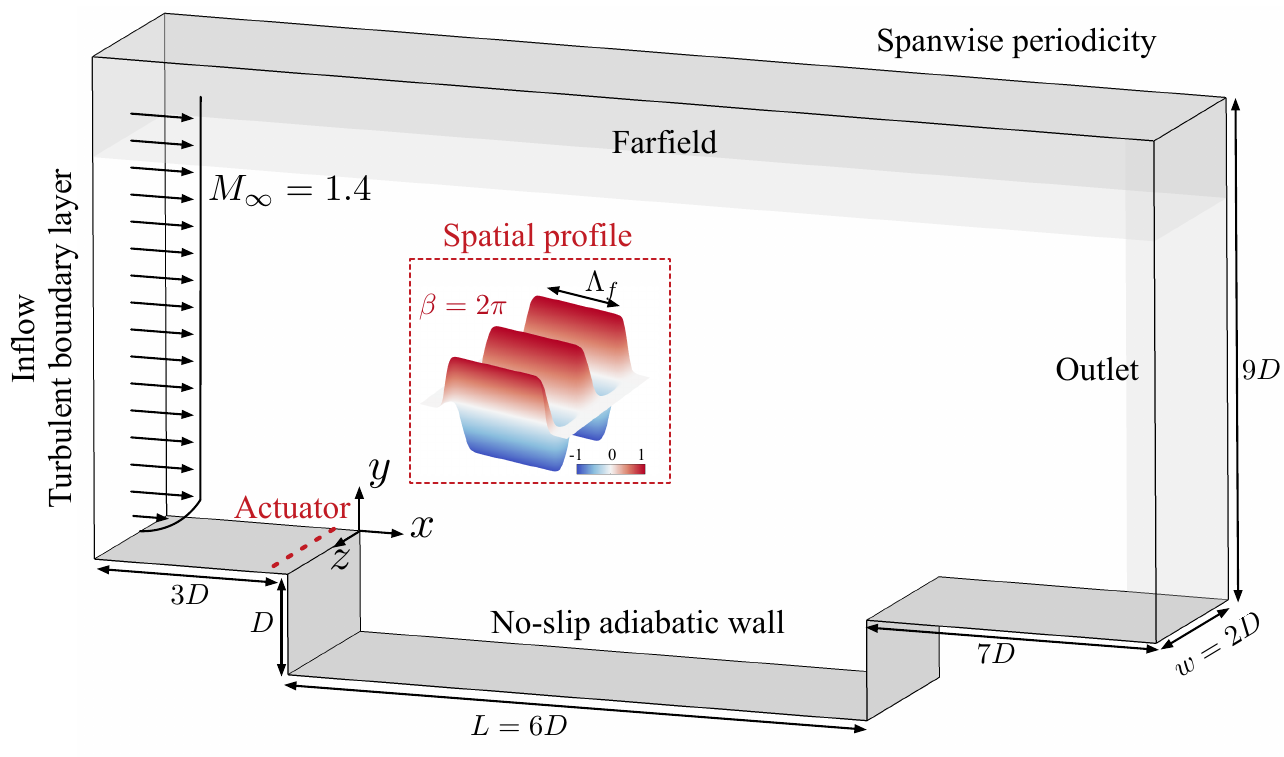}
    \caption{Computational setup (not to scale) with spatial profile of actuator for $\beta=2\pi$}
    \label{fig:domain}
\end{figure}

\subsection{Computational setup}
We perform the large eddy simulation (LES) of the turbulent cavity flow using the compressible flow solver CharLES, which uses a second-order finite volume spatial discretization scheme and a third-order accurate Runge-Kutta scheme for time stepping \citep{ham2004energy,ham2006accurate}. The Vreman model is used for sub-grid scale modelling \citep{vreman2004eddy}, and Harten-Lax-van Leer scheme \citep{toro1994restoration} is used as the shock capturing scheme. The computational domain (not to scale) is shown in figure~\ref{fig:domain}. The incoming streamwise velocity profile in the boundary layer uses a one-seventh power law that smoothly joins with the free stream, superimposed with random Fourier modes to simulate unsteady fluctuations at the inlet \citep{bechara1994stochastic}. The initial boundary layer thickness at the leading edge is set to be $\delta_0/D=0.167.$ The far-field and outlet boundary conditions are specified as sponge zones to dampen the exiting waves and prevent any numerical reflections. No-slip and adiabatic boundary conditions are specified along the cavity walls. Periodicity is enforced in the spanwise direction.

The origin of the Cartesian coordinate system is set at one of the spanwise boundaries, as shown in figure~\ref{fig:domain}. The $x,\,y$ and $z$ coordinates denote the streamwise, transverse and spanwise directions. The spanwise periodic extent is two times the depth of the cavity. The inlet is located at three depths from the leading edge of the cavity, and the outlet is specified at seven depths from the aft cavity wall. The farfield boundary condition is specified at nine depths in the transverse direction from the cavity leading edge. We perform local refinement near the actuator at the leading edge location in all three directions. The computational setup is the same as the one used in our earlier numerical and experimental studies and has been verified extensively \citep{zhang2019suppression, sun2019effects, liu2021unsteady}. 

\subsection{Actuator setup}

We compute impulse response and perform flow control of cavity flow using a wall-normal unsteady blowing jet near the leading edge of the cavity. The actuator setup is chosen to match our previous numerical and experimental studies \citep{zhang2019suppression, liu2021unsteady,singh2021uncovering}. The streamwise location of the actuator is set at $x_f/D=-0.0698$, and the streamwise width of the actuator slot is $\Lambda_f/D=0.0175$. The unsteady blowing at a spanwise wavenumber, $\beta$ is prescribed by modifying the wall-normal velocity boundary condition as 
\begin{equation}
    u_{\textrm{jet}}(x,z,t)=\eta(t)h(x,x_f,\Lambda_f)\cos(\beta z),
\end{equation}
where $\eta(t)$ is a temporal waveform of the unsteady actuation, $h(x,x_f,\Lambda_f)$ is the streamwise shape function of the actuation, and the last term accounts for the spanwise sinusoidal behavior. The streamwise spatial profile of the actuator is specified using a hyperbolic tangent function to prevent numerical discontinuity near the actuator slot edges as,
\begin{equation}
    h(x,x_f,\Lambda_f)=\frac{1}{4}\left[1+\tanh(\kappa_1(x-x_f+\Lambda_f/\kappa_2))\right]\left[1-\tanh(\kappa_1(x-x_f-\Lambda_f/\kappa_2))\right],
\end{equation}
where $\kappa_1=2000$ and $\kappa_2=2.6$. The spatial profile of the actuator is shown in figure~\ref{fig:domain}.

We perturb the cavity flow using a Gaussian impulse to compute the phase sensitivity function, as will be discussed in later sections. The temporal waveform of the impulse is 
\begin{equation}
    \eta(t) = u_{\textrm{impulse}}(t)= \frac{\epsilon}{\sqrt{2\pi}\sigma} \exp\left[-\frac{1}{2}\frac{(t-t_0)^2}{\sigma^2}\right],
    \label{eq:impulse_pert}
\end{equation}
where $\epsilon$ is the magnitude of impulse perturbation, $t_0$ is the time-instant of the impulse perturbation, $\sigma$ is the pulse-width of the impulse. For the current study, we consider an impulse perturbation of magnitude $\epsilon=0.05$, and $\sigma$ is chosen to be one percent of the dominant oscillation period. 

We consider open-loop flow control using two periodic waveforms with an actuation frequency $\omega_f=2\pi f_c$ and a Strouhal number $St_L^c=f_cL/U_\infty$. One is a periodic non-sinusoidal phase-sensitivity-based actuation waveform $\eta_{Z}(\omega_f t)$ (discussed later in Sec. 4.3) and the other is a sinusoidal actuation waveform at the same frequency $\eta_{\textrm{sine}}(t)=A\sin(\omega_f t)$, where $A$ is the sinusoidal actuation amplitude. 

The actuation cost is measured using the unsteady momentum coefficient, $C_\mu$. The unsteady momentum coefficient for active flow control cases is defined as
\begin{equation}
    C_\mu = \frac{1}{1/2 U_\infty^2 w\delta_0}\left[\frac{\omega_f}{2\pi}\int_{T_f}\int_{l_x}\int_{l_z} u_{\textrm{jet}}(x,z,\omega_f t)^2 \,dz\,dx\,dt\right],
\end{equation}
where $T_f=2\pi/\omega_f$ is the actuation time-period, $l_x$ and $l_z$. are the spanwise extents of actuation in $x$ and $z$ directions, respectively and $w=2D$ is the spanwise extent of the cavity. In this study, we consider $ C_\mu=0.02$ in line with previous studies \citep{williams2007supersonic,zhang2019suppression,liu2021unsteady} to assess the control performance.

Since the control objective in this study is to reduce the pressure fluctuations within the cavity, the control performance is measured using relative change of overall root mean squared (r.m.s.) values of pressure distribution between the controlled $(\tilde{p}_{\textrm{rms}}^{\textrm{c}})$ and uncontrolled flows $(\tilde{p}_{\textrm{rms}}^{\textrm{unc}})$ in the cavity as
\begin{equation}
    \Delta \tilde{p}_{\textrm{rms}}=\frac{\tilde{p}_{\textrm{rms}}^{\textrm{unc}}- \tilde{p}_{\textrm{rms}}^{\textrm{c}}}{\tilde{p}_{\textrm{rms}}^{\textrm{unc}}}\quad \text{and} \quad \tilde{p}_{\textrm{rms}}=\int_\mathcal{V}\frac{p_{\textrm{rms}}}{\frac{1}{2}\rho U_\infty^2} dV,
\end{equation}
where $\tilde{p}_{\textrm{rms}}^{}$ is volume-integrated pressure fluctuations and $\mathcal{V}$ is the domain of integration, $(x,\,y,\,z)/D \in ([0,6],[-1,1.5],[0,2])$.

 \begin{figure}
\centering
    \includegraphics[width=\textwidth]{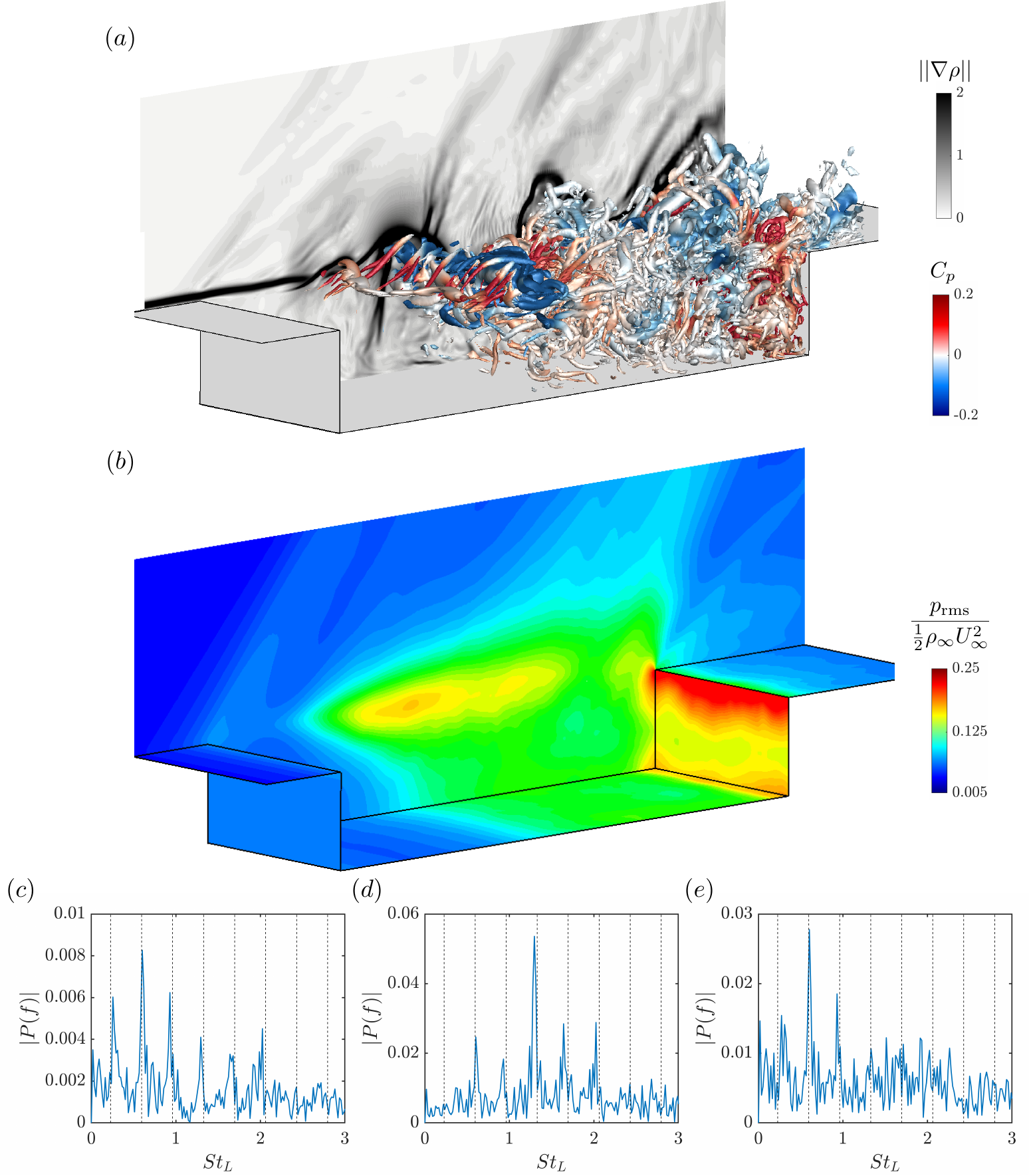}
    \caption{(a) Instantaneous flowfield visualized using iso-surface of $Q-$criterion=14 and colored by the coefficient of pressure. The background shows numerical schlieren in grey. (b) The normalized r.m.s. pressure along the cavity walls and a slice at a spanwise plane $z=0$. (c)-(e) Frequency spectra $|P(f)|$ of the pressure probes in the shear layer located at $(x,\,y,\,z)/D=(0.75,0,1),\,(3,0,1)$ and $(5.25,0,1)$, respectively, with superposed Rossiter mode frequencies denoted by the vertical dotted lines.}
    \label{fig:baseflow}
\end{figure}

 \section{Uncontrolled cavity flow}
 \label{sec:methodology}
 \subsection{Flow physics}

Let us consider supersonic turbulent flow over an open rectangular cavity at $M_\infty=1.4$ and $Re=10000$. The representative instantaneous flowfield visualized using an iso-surface of $Q$-criterion and colored by the coefficient of pressure is shown in figure~\ref{fig:baseflow}(a). The numerical schlieren visualized using $||\nabla \rho||$ for a spanwise plane $z=0$ is shown in the background. The shear layer emanating from the leading edge rolls up, forming large-scale spanwise vortical structures at one-third of the cavity length from the leading edge. Since the flow is supersonic, we also observe compression waves generated due to these vortical structures obstructing the freestream flow. Past mid-cavity, we observe the emergence of small-scale structures around the large-scale structures. These large structures convect downstream and impinge on the aft wall, generating strong acoustic waves that can travel upstream through the subsonic regime within the cavity and perturb the leading edge shear layer, creating a feedback loop. The vortical structures and the compression waves produce violent fluctuations inside the cavity. 

The distribution of normalized r.m.s. pressure $p_{\textrm{rms}}/(\frac{1}{2}\rho_\infty u_\infty^2)$ along the cavity walls and on a plane $z=0$ is visualized in figure~\ref{fig:baseflow}(b). We observe large fluctuations on the aft wall of the cavity due to the impingement of the vortical structures. Further, we also observe a secondary region with high r.m.s. of pressure fluctuations in the mid-cavity due to the convection of the formed of large-scale spanwise vortical structures.

This complex nonlinear flow physics is comprised of various time scales. The resonant acoustic tones in the cavity that occur due to the acoustic feedback are referred to as the Rossiter modes. These resonant frequencies in the cavity can be determined based on the incoming free-stream Mach number using Rossiter's semi-empirical formula \citep{heller1971flow},
\begin{equation}
    St_L^n  = \frac{n-\alpha}{1/k+M_\infty/\sqrt{1+(\gamma-1)M_\infty^2/2}},
\end{equation}
where $St_L^n=fL/U_\infty$ is the length-based Strohaul number for $n$-th Rossiter mode, $\gamma=1.4$ is the specific heat ratio, $k=0.65$ is the averaged convection speed of disturbance and $\alpha=0.38$ is the phase-delay Rossiter mode \citep{liu2021unsteady}. 

The frequency spectra $|P(f)|$ of three pressure probes in the shear layer at the locations $(x,\,y,\,z)/D=(0.75,0,1),\, (3.0,0,1),\,(5.25,0,1)$ are shown in figure~\ref{fig:baseflow}(c)-(e). The black dotted lines indicate the predicted first eight Rossiter modes. The dominant time scales match well with the predicted frequencies and earlier studies \citep{liu2021unsteady}. In the shear layer closer to the leading edge, the second Rossiter mode $St_L=0.6$ is dominant, and the first and third Rossiter modes, $St_L=0.23$ and $0.96$, are sub-dominant. Moving towards the mid-cavity, we observe a strong peak at the fourth Rossiter mode $St_L=1.32$. The dominant frequency shifts to the lower second Rossiter mode near the aft wall. We observe that the maximum oscillation energy is observed in the mid-cavity region due to the formation and convection of the large-scale spanwise structures.

\begin{figure}
\centering
    \includegraphics[width=0.8\textwidth]{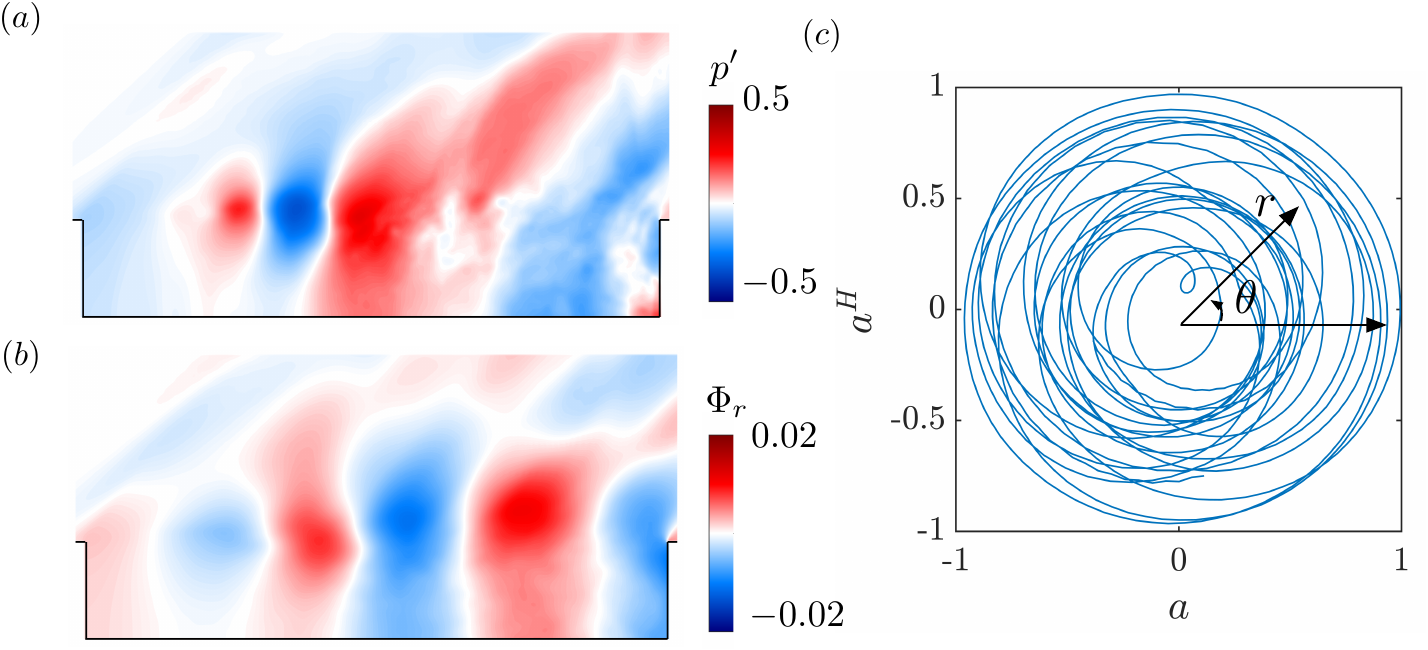}
    \caption{Phase description for turbulent cavity flow. (a) Instantaneous spanwise-averaged pressure fluctuation $p^\prime$. (b) The real component of the DMD mode of spanwise-averaged pressure at $St_L\approx1.32$. (c) phase space $a$-$a^H$ obtained using the inner product of the pressure fluctuations with the real component of the DMD mode.}
    \label{fig:phase_description}
\end{figure}

\begin{figure}
\centering
    \includegraphics[width=\textwidth]{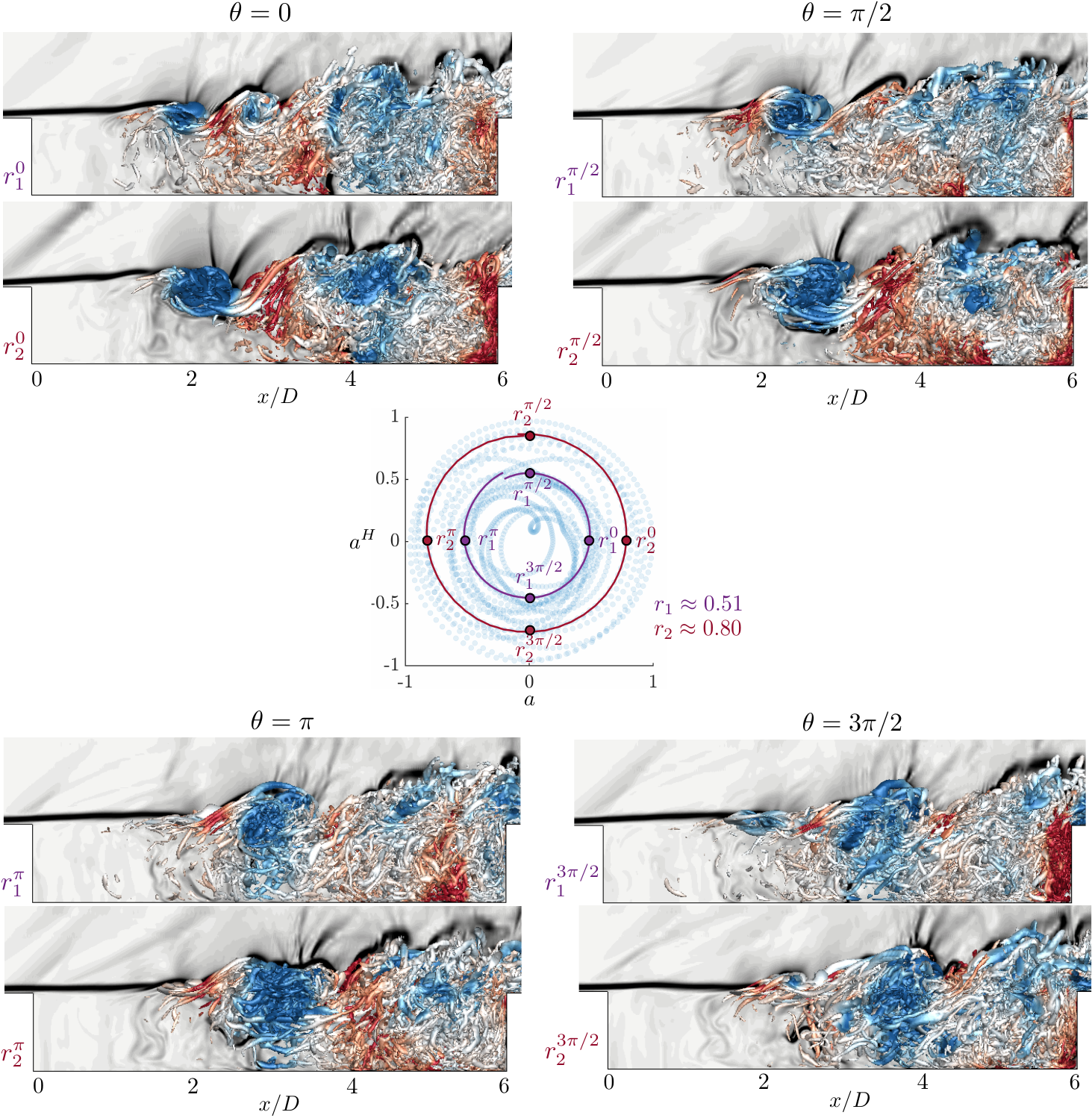}
    \caption{Instantaneous flowfields visualized using iso-surface of $Q-$criterion=14 and colored by the coefficient of pressure corresponding to various phases $\theta=[0,\pi/2,\pi,3\pi/2]$ and radii, $r\approx0.51$ and $0.8$ respectively in the phase space.}
    \label{fig:phase_dynamics}
\end{figure}

 \subsection{Definition of phase}
 
The dominant features of the cavity flows are characterized by the various phases of formation, convection and the impingement of the large vortical structures in the cavity. We aim to identify low-dimensional phase dynamics that can effectively capture these various phases of vortex convection in the cavity. To this end, we need to identify the dominant time scale and spatial structures associated with the convection process. From the frequency spectra of the pressure probes shown in figure~\ref{fig:baseflow}(c)-(e), we observe broadband spectra with various dominant frequencies depending on the spatial location. Since the large-vortex roll-up occurs about the mid-cavity, the dominant Rossiter mode associated with the convection is $St_L=1.32$ based on the mid-cavity pressure probe (see figure~\ref{fig:baseflow}(d)). 

To support this analysis, we leverage modal analysis \citep{taira2017modal} to identify a low-dimensional space that can capture the dominant physics. More specifically, we apply dynamic mode decomposition (DMD) \citep{rowley2009spectral, schmid2010dynamic,tu2013dynamic} to identify the spatial structure corresponding to the vortex convection at the corresponding dominant frequency. Since we are mainly interested in capturing the large vortical structures, the DMD is performed on the spanwise-averaged pressure field. The DMD is performed using 30 cycles associated with the dominant frequency, $\omega_d$ using about 80 snapshots per period. The instantaneous spanwise-averaged pressure fluctuations ($p^\prime (x,y,t)$) and the real component of the DMD mode $\Phi_r(x,y)$ corresponding to $St_L\approx 1.32$ are shown in figure~\ref{fig:phase_description}(a) and (b), respectively. The DMD mode highlights the shear-layer region around the mid-cavity. 

We can capture the relative phases of the convective process of these large-scale structures by projecting the instantaneous pressure field onto the DMD mode. Since the DMD mode is complex and the real and imaginary parts are shifted by a phase, we consider the inner product ($\langle \cdot , \cdot \rangle$) of the instantaneous pressure fields with the real component of the DMD mode given by
\begin{equation}
    a(t) \equiv \langle p^\prime(x,y,t), \Phi_r(x,y) \rangle.
    \label{eq:DMD_coeff}
\end{equation}

We now establish a two-dimensional phase space that can capture the dominant flow physics by defining the phase $\theta$ and amplitude $r$ based on the time signal $a(t)$. While several ways exist to compute the phase and amplitude from a temporal signal, we consider the widely used Hilbert transform \citep{boashash1992estimating, Pikovsky_Rosenblum_Kurths_2001} that also results in removing high frequency fluctuations. Using the Hilbert transform of $a(t)$ denoted by $a^H(t)$, the phase and amplitude can be defined in the normalized $a-a^H$ phase space as
\begin{align}
    \theta(t)\equiv\tan^{-1}\left(\frac{a^H(t)}{a(t)}\right),\quad r(t)\equiv\sqrt{a(t)^2+{a^H}(t)^2}.
    \label{eq:phase_hilbert}
\end{align}
Here, we note that since the DMD modes are unique up to an arbitrary phase shift, the phase variable $\theta$ discussed in this study could be shifted, and absolute values are not unique, but the relative phase dynamics are of importance. We maintain consistency of $\theta$ using the same DMD mode for all the analysis. We also note that the phase obtained depends on the method used to compute the instantaneous phase. For instance, in addition to the Hilbert transform, one can also consider $a(t)$ and $\dot{a}(t)$ for obtaining a phase space which results in a phase shift of $\theta(t)$ relative to that obtained using the Hilbert transform. The comparison between these methods is discussed in the Appendix.

The phase space $a-a^H$ for the spanwise-averaged cavity flow is shown in figure~\ref{fig:phase_description}(c). Since turbulent cavity flow exhibits dynamics with broadband spectra, we observe a noisy limit cycle due to the cycle-to-cycle variation in the amplitude and the time period with an averaged frequency of $St_L=1.32$. The radius $r$ characterizes the pressure fluctuations within the cavity at this frequency, as it is based on the magnitude of the inner product with the DMD mode. The phase $\theta$ captures the various stages of vortex convection within the cavity. The correlation between the phase variable and the flow physics is depicted in figure~\ref{fig:phase_dynamics}. To identify the dependence on radius $r$, we compare the flow physics at two arbitrary radii as a demonstration. For instance, the $xy$-view of instantaneous flowfields corresponding to two average radii, $r_1\approx 0.51$ and $r_2\approx 0.8$ (corresponding cycles are highlighted in the phase space figure~\ref{fig:phase_dynamics}(middle)) at four different phases are shown in figure~\ref{fig:phase_description}.

For all the phases, the flowfields corresponding to the smaller radius, $r_1\approx 0.51$ exhibit weaker vortical structures with lower fluctuations within the cavity, whereas the cycle corresponding to the larger radius shows the presence of stronger spanwise vortical structures within the cavity. This can also be observed by the strong compression waves seen in the background schlieren for instantaneous flowfields corresponding to $r_2\approx 0.8$. Now, let us consider the evolution of flowfields relative to the phase $\theta$. For both radii $r$, when $\theta=0$ (top left), we observe the presence of a strong spanwise vortex closer to $x=2D$. We then observe the convection of this strong vortex to mid-cavity at $\theta=\pi/2$ (top right). As we move further along $\theta$, at phase $\theta=\pi$ (bottom left), we observe that the strong large-scale structure diffuses, and we start to see the emergence of several small-scale structures around this vortex, about the mid-cavity $x=3D$. Further, at the phase $\theta=3\pi/2$ (bottom right), we observe the loss of coherence of the large-scale vortex and the presence of several small-scale structures from $x/D=3.5$ to 4. At this phase, we also see the start of the shear-layer roll-up upstream of $x/D=2$ that precedes the spanwise vortex at $\theta=0$, forming an oscillation cycle. The relative position of the large-scale structure to the cavity length is captured well by the phase variable defined by the frequency of $St_L=1.32$. Thus, by projecting the instantaneous flowfield onto the DMD mode, we can capture the cycle of vortex convection using the phase variable.

 \section{Phase-sensitivity-based actuation waveform}
\label{sec:phase_reduction}
 
With the phase dynamics describing the large-scale vortex convection, we can leverage phase reduction analysis to find an actuation waveform optimized for quick flow modification. While phase reduction analysis has been traditionally used to analyze limit cycles, we extend its application to multi-modal turbulent flows. We first identify the phase sensitivity function, which identifies the optimal timing that leads to maximal phase response, and then analytically obtain an actuation waveform by solving an optimization problem. 

 \subsection{Phase reduction approach for periodic flows}

Let us consider compressible periodic fluid flows governed by the Navier--Stokes equations, 
\begin{equation}
\dot{\boldsymbol{q}}=\mathcal{N}(\boldsymbol{q}(\boldsymbol{x},t)),
\label{eq:governing}
\end{equation} 
where $\boldsymbol{q}=[\rho,\,\rho u_x,\,\rho u_y,\,\rho u_z,\,e]$ and $\mathcal{N}$ is the nonlinear Navier--Stokes operator. Here, $\rho, u_x,\, u_y,\, u_z,\,e$ are the density, velocities in streamwise, transverse and spanwise directions and energy, respectively. 

The periodic solution, $\boldsymbol{q}_0(\boldsymbol{x},t)$ represents a limit cycle and satisfies $\boldsymbol{q}_0(\boldsymbol{x},t)=\boldsymbol{q}_0(\boldsymbol{x},t+T)$, where $T$ is the time-period of the oscillation with oscillation frequency $\omega_n=2\pi/T$. For such a flow, the phase $\theta\in [0,2\pi)$ defined on the limit cycle satisfies
\begin{equation}
    \dot{\theta}(t) = \omega_n.
    \label{eq:phase_periodic}
\end{equation}
Given this perspective, we can map the flow state on the limit cycle $\boldsymbol{q}_0(\boldsymbol{x},\theta)$ to the single scalar phase variable $\theta$. For periodic flows, the phase variable $\theta$ can be defined on the phase-space formed by characteristic observable of the fluid system. For instance, the lift coefficent has been used to define $\theta$ for periodic wake flows \citep{taira2018phase}. In this work, we consider the phase space defined based on the global DMD mode for turbulent oscillatory flows as shown in figure~\ref{fig:phase_description} and the corresponding phase description is discussed in section~\ref{subsec:method}.

If we have a stable limit cycle solution, this concept of phase can be extended to the vicinity of the limit cycle solution through a generalized phase variable $\Theta(\boldsymbol{q}(\boldsymbol{x},t))$ such that $\Theta(\boldsymbol{q})=\theta$. Thus combining equations~\ref{eq:governing} and~\ref{eq:phase_periodic} and following chain rule, the generalized phase variable satisfies
\begin{equation}
     \dot{\theta}=\dot{\Theta}(\boldsymbol{q}) = \int_{\mathcal{D}}\nabla_{\boldsymbol{q}} \Theta(\boldsymbol{q}) \cdot \dot{\boldsymbol{q}} d \boldsymbol{x} = \int_{\mathcal{D}}\nabla_{\boldsymbol{q}}\Theta(\boldsymbol{q}) \cdot \mathcal{N}(\boldsymbol{q}) d \boldsymbol{x} = \omega_n,
\end{equation}
where $\mathcal{D}$ corresponds to the fluid region.

We can now obtain the phase response to sufficiently small perturbations.
We consider the perturbation of the form $\epsilon\boldsymbol{p}(\boldsymbol{x},t)$ added to the governing equations as,
\begin{equation}
    \dot{\boldsymbol{q}}=\mathcal{N}(\boldsymbol{q}(\boldsymbol{x},t))+\epsilon\boldsymbol{p}(\boldsymbol{x},t)
\end{equation}
The phase dynamics can be derived as,
\begin{align}
   \dot{\theta}(t)&=\dot{\Theta}(\boldsymbol{q}) = \int_{\mathcal{D}} \nabla_{\boldsymbol{q}}\Theta(\boldsymbol{q}) \cdot \dot{\boldsymbol{q}} d \boldsymbol{x} = \int_{\mathcal{D}} \nabla_{\boldsymbol{q}} \Theta(\boldsymbol{q}) \cdot \left[\mathcal{N}(\boldsymbol{q}(\boldsymbol{x},t))+\epsilon \boldsymbol{p}(\boldsymbol{x},t)\right] d \boldsymbol{x} \nonumber\\
&\approx  \omega_n + \epsilon\int_{\mathcal{D}}\boldsymbol{Y}(\boldsymbol{x},\theta)\cdot \boldsymbol{p}(\boldsymbol{x},t) d \boldsymbol{x},
\label{eq:phase_dyn} 
\end{align}
where $\boldsymbol{Y}(\boldsymbol{x},\theta)\equiv\nabla_{\boldsymbol{q}} \Theta(\boldsymbol{q})|_{\boldsymbol{q}=\boldsymbol{q}_0(\boldsymbol{x},\theta)}$ is the first-order spatial phase sensitivity function. It quantifies the linear phase response, i.e., phase advancement or delay for any arbitrary small perturbation. We can further simplify the equation \ref{eq:phase_dyn}, by considering a perturbation of the form $\boldsymbol{p}(\boldsymbol{x},t)=\boldsymbol{h}(\boldsymbol{x})\eta(t)$ such that,
\begin{align}
    \dot{\theta}(t)=\omega_n + \epsilon\int_{\mathcal{D}}\boldsymbol{Y}(\boldsymbol{x},\theta)\cdot \boldsymbol{p}(\boldsymbol{x},t) d \boldsymbol{x} = \omega_n + \epsilon Z(\theta)\eta(t),
    \label{eq:effective_phase}
\end{align}
where $Z(\theta)=\int_{\mathcal{D}}\boldsymbol{Y}(\boldsymbol{x},\theta)\cdot \boldsymbol{h}(\boldsymbol{x}) d \boldsymbol{x}$ is the effective phase sensitivity function. 

There are various ways to compute the phase sensitivity function for periodic fluid flows: the adjoint method \citep{kawamura2013collective,kawamura2015phase,kawamura2022adjoint}, the direct method \citep{taira2018phase,khodkar2020phase,nair2021phase,loe2021phase} and the Jacobian-free approach \citep{iima2021phase}. In this study, we use an extension of the direct impulse-based method to compute phase sensitivity, as this method can be used in both simulations and experiments. The direct impulse-based method for periodic flows is based on evaluating the phase shift at various phases in response to added impulse perturbations. Once the phase space and phase variables are determined, at a phase, $\theta_0=\theta(t_0)$, a small impulse perturbation of the form $\epsilon\eta(t)=\epsilon\delta(t-t_0)$, where $\delta(t)$ is the Dirac-delta distribution, is added to the periodic flow. The phase of the perturbed flow, $\theta_{\textrm{pert}}$ is then from the phase space of the perturbed flow. For periodic flow, this small impulse creates a phase shift once the transient effects of the perturbation decay. This shift can be measured using the phase difference between the perturbed and the unperturbed flows as, $g(\theta_0;\epsilon)=\lim_{t\rightarrow \infty}\theta_{\textrm{pert}}(t)-\theta_{\textrm{unpert}}(t)$ \citep{taira2018phase}. Thus the first-order phase response is given by $Z(\theta_0)=g(\theta_0; \epsilon)/\epsilon$.

\subsection{Phase response for turbulent oscillatory flows} 
\label{subsec:method}
\begin{figure}
    \centering
    \includegraphics[width=0.93\textwidth]{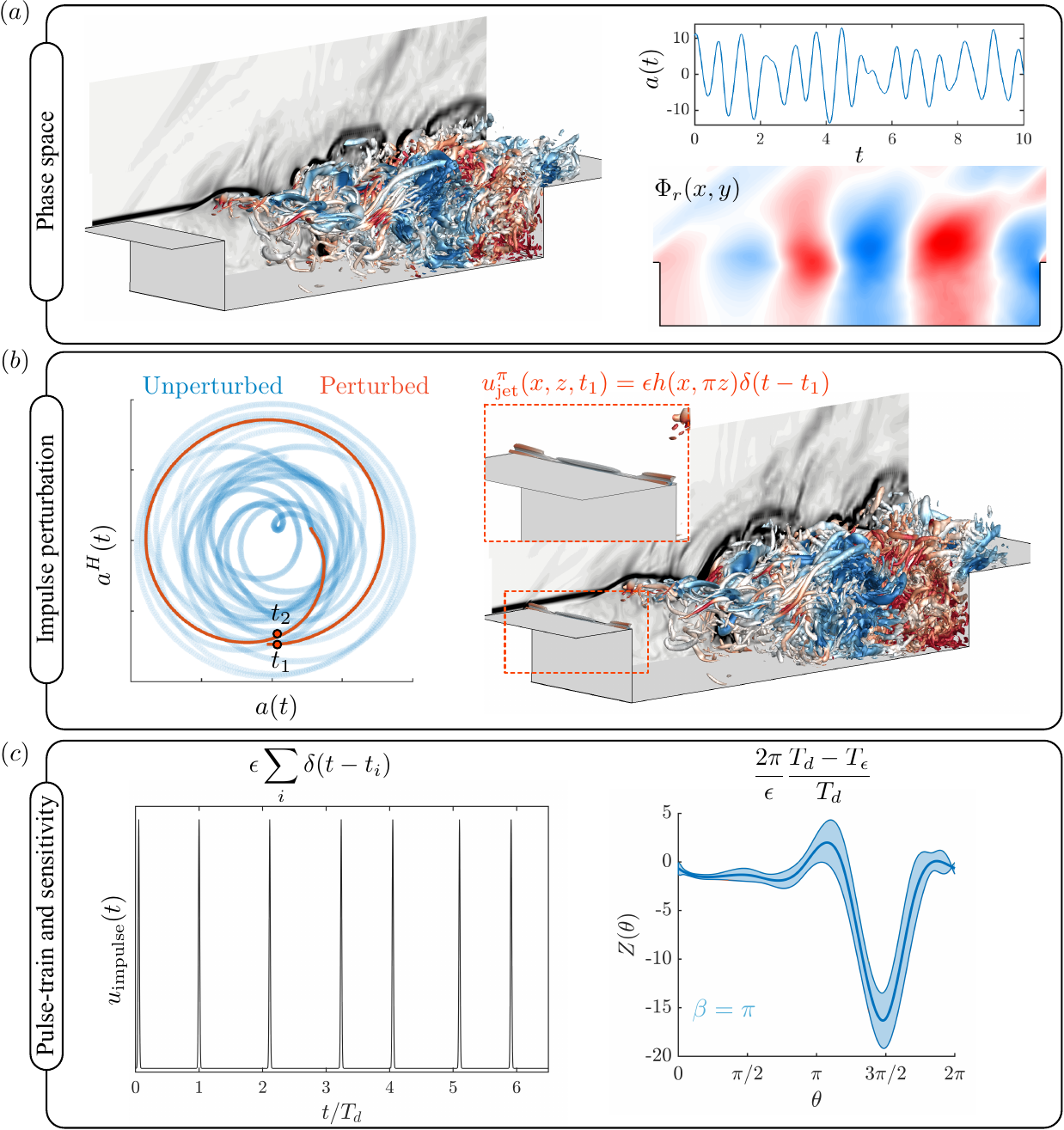}
    \caption{Overview of continuous pulse method to compute phase sensitivity of chaotic flows. (a) Definition of phase space using DMD mode of the dominant frequency $St_L=1.32$. (b) First instant of impulse perturbation added at the leading edge with a spanwise wavenumber $\beta=\pi$, when $\theta_0=3\pi/2$ and the corresponding perturbed trajectory in the $a-a^H$ phase space. (c) Pulse train perturbation $u_{\textrm{impulse}(t)}$ and phase sensitivity function $Z(\theta)$ measured using averaged time-shift between perturbed and unperturbed trajectories.}
    \label{fig:overview}
\end{figure}

Traditionally, the phase reduction analysis is applied to periodic systems. Since many high-speed fluid flows are turbulent, exhibiting a dominant oscillatory behavior, we extend the phase reduction approach for such complex oscillatory flows. The computation of phase response of such flows is challenging due to a couple of factors. Firstly, due to the chaotic dynamics (resulting in the absence of strict limit cycles in phase space), there is no asymptotic phase shift as measured using the direct impulse-based method. Further, due to the broadband nature of turbulent flows superimposed on the dominant frequency, there is cycle-to-cycle variation in frequency and time period. The overview of phase reduction for turbulent oscillatory flows is shown in figure~\ref{fig:overview}.

Given these challenges, there have been a limited number of studies that provide heuristic approaches to compute phase sensitivity functions for low-dimensional chaotic oscillators \citep{josic2001phase,schwabedal2012optimal,kurebayashi2012theory,imai2022phase,tonjes2022phase}. Among these methods, the approach proposed by \cite{tonjes2022phase} facilitates a direct extension of the impulse-based method to compute the phase sensitivity function. Their approach measures the averaged phase response behaviour of the chaotic system relative to the average oscillatory period of the unperturbed system. We implement their approach to compute the averaged phase sensitivity function for turbulent cavity flow.


Following the phase dynamics for periodic flows in equation \ref{eq:effective_phase}, the phase dynamics of perturbed chaotic flows can be described as 
\begin{align}
    \dot{\theta} = \omega_d+D_\omega(t) + \epsilon Z(\theta)\eta(t)
    \label{eqn:chaotic_phase_dyn_v1}
\end{align}
where $\omega_d$ is the dominant frequency in the phase space and $D_\omega(t)$ denotes the fluctuations about the averaged dominant frequency due to the chaotic flow dynamics. Note that, $\int\limits_0^\infty D_\omega(t)\,dt =0$.

Extending the direct impulse-based method, we consider a continuous pulse train perturbation whenever a trajectory crosses a phase $\theta_0$ \citep{tonjes2022phase}. We consider the perturbation of the form, $\eta(t)=\sum_i\delta(t-t_i)$ (approximated using Gaussian impulse profile in equation~\ref{eq:impulse_pert}), where $t_i$ is the time instant corresponding to phase $\theta(t_i)=\theta_0$ of the perturbed trajectory. The phase dynamics of continuous $N$-pulse perturbed chaotic flows is given by,
\begin{align}
    \dot{\theta} = \omega_d+D_\omega(t) + \epsilon \sum\limits_{i=1}^N Z(\theta)\delta(t-t_i).
    \label{eqn:chaotic_phase_pulse}
\end{align}
For sufficiently large $N$, the averaged phase dynamics about the unperturbed system is given as,
\begin{align}
   \langle \dot{\theta} \rangle_{t} = \omega_d + \epsilon \sum\limits_{i=1}^N \langle Z(\theta)\delta(t-t_i) \rangle_{t},
    \label{eqn:chaotic_phase_averaged}
\end{align}
where $\langle \cdot \rangle_t$ is temporal average. Thus, the averaged phase sensitivity can be computed based on the shift in the average time period and can be written in terms of the average periods of perturbed, $T_\epsilon$ and unperturbed, $T_d$ as
\begin{align}
    Z(\theta) = \frac{2\pi}{\epsilon}\frac{T_d - T_\epsilon}{T_d}.
    \label{eq:phase_sensitivity_turbulent}
\end{align}

The overall procedure of computing the phase sensitivity function for turbulent oscillatory flows is as follows and is shown in figure~\ref{fig:overview}:
\begin{itemize}
    \item \textbf{Defining phase variable $\theta$}: We define the phase variable $\theta$ using the inner product between spanwise averaged pressure field and the DMD mode associated with $St_L=1.32$ as $a(t)=\langle p^\prime(x,y,t),\Phi_r(x,y)\rangle$ (see figure~\ref{fig:overview}(a)). The phase and radius variables are defined using the $a-a^H$ phase space (unperturbed trajectory in figure~\ref{fig:overview}(b)) using equation~\ref{eq:phase_hilbert}.
    \item \textbf{Addition of impulse perturbation} (figure~\ref{fig:overview}(b)): To measure the response of the system at each phase, we introduce an impulse perturbation at the leading edge at a spanwise wavenumber $\beta$. An impulse perturbation is added to the unperturbed flow at a time instant $t_1$ corresponding to a phase $\theta_0$. An instance of this process for $\beta=\pi$ and $\theta_0=3\pi/2$ is shown in figure~\ref{fig:overview}(b).  The profile of this perturbation following equation~\ref{eq:impulse_pert} with a zoomed-in view at the actuator location can be seen for $\beta=\pi$. The simulation is then evolved with the added impulse and the perturbed trajectory is then obtained as $a_{\textrm{pert}}(t) =\langle p_{\textrm{pert}}^\prime(x,y,t),\Phi_r(x,y)\rangle$, where $p_{\textrm{pert}}^\prime$ is the spanwise-averaged pressure field for the perturbed cavity flow. The phase variable $\theta_{\textrm{pert}}(t)$ is measured using $a_{\textrm{pert}}-a^H_{\textrm{pert}}$ phase space and an instance of this perturbed trajectory is shown in red in figure~\ref{fig:overview}(b). The second impulse is introduced at the next time instant $(t_2)$ when the perturbed trajectory crosses $\theta_{\textrm{pert}}(t_2)=\theta_0$.
    \item \textbf{Computing phase sensitivity through pulse-train method} (figure~\ref{fig:overview}(c)):  The impulse addition is repeated for $N$ cycles for each $\theta_0$ by at $t_i,i=1,2,\cdots,N$ such that $\theta_{\textrm{pert}}(t_i)=\theta_0$. An example of a pulse-train perturbation, $u_{\textrm{impulse}}(t)$ is shown in (figure~\ref{fig:overview}(c)). The time instants of the impulses are plotted relative to the dominant averaged period, $T_d=2\pi/\omega_d$, to visualize the cycle-to-cycle variations of the trajectory. The averaged phase sensitivity $Z(\theta)$ is then measured as the cumulative shift in the time period of the perturbed trajectory relative to the unperturbed trajectory (equation~\ref{eq:phase_sensitivity_turbulent}). For practical applications of this method, we use a finite number of pulses, $N$, and the converged phase sensitivity relative to $N$ is considered for further analyses. The phase sensitivity function $Z(\theta)$  after 13 cycles is shown in figure~\ref{fig:overview}(c).
\end{itemize}

\subsection{Optimal actuation waveform for fast synchronization}

While the flow control objective for cavity flows is aimed at suppressing the overall pressure fluctuations, it is also desirable for the control to take effect and suppress the fluctuations as quickly as possible to minimize structural fatigue. Based on the averaged phase sensitivity function, we now consider an optimal actuation waveform that quickly modifies the dominant frequency of the vortex convection dynamics in the shear layer \citep{zlotnik2013optimal,godavarthi2023optimal}. We achieve this by considering a waveform that can quickly synchronize to the forcing frequency. While lock-on/synchronization is undesirable for cavity flows, the phase-based analysis identifies the optimal waveform in a linear sense. Since the underlying flowfield has several other dominant tones and nonlinearities, we hypothesize that the obtained waveform with sufficiently small amplitude at a detuned frequency will suppress the dominant cavity tone. The large vortical structures in the cavity responsible for the violent fluctuations arise due to the acoustic feedback. We aim to quickly modify the frequency of the vortex convection to disrupt this feedback and break down the large-scale structures, thereby reducing the overall pressure fluctuations within the cavity. To this end, we obtain an actuation waveform for fast frequency modification through the lens of synchronization.

 Following equation~\ref{eqn:chaotic_phase_averaged}, synchronization to an external periodic force with a forcing frequency, $\omega_f$ can be analyzed using the relative phase dynamics, $\psi(t)=\theta(t)-\omega_f t$ as
 \begin{equation}
\dot{\psi}(t) = \Delta \omega + \epsilon Z(\psi+\omega_f t)\eta(\omega_f t),
 \end{equation}
 where $\Delta \omega = \omega_d-\omega_f$. Synchronization is achieved when $\dot{\psi}(t) \rightarrow 0$. Since this equation has explicit temporal dependence, we can transform it into an autonomous form by averaging over a period of forcing \citep{kuramoto1984chemical, ermentrout1991multiple} to obtain,
 \begin{equation}
     \dot{\psi}(t) = \Delta \omega + \epsilon\Gamma(\psi),\quad \Gamma(\psi)=\frac{1}{2\pi}\int\limits_{0}^{2\pi} Z(\psi+\phi)\eta(\phi)\, d\phi,
      \label{eq:relative_phase}
 \end{equation}
 where $\Gamma(\psi)$ is the phase-coupling function. Therefore, the synchronization condition to an external forcing can be theoretically identified as $\epsilon \min\Gamma(\psi) \leq -\Delta \omega \leq  \epsilon  \max\Gamma(\psi)$ \citep{nakao2016phase}.

Now, we seek an optimal actuation waveform to achieve quick frequency modification by maximizing $|\dot{\psi}_*|$, where $\psi_*$ is the stable fixed point of the relative phase dynamics. The cost function for minimization can then be expressed as  
\begin{equation}
    \mathcal{J}(\eta) = -\Gamma^\prime(\psi_*) - \lambda\left( \langle\eta^2\rangle - 1 \right) - \mu\left[\Delta \omega+\epsilon\Gamma(\psi_*)\right],
    \label{eq:optimization}
\end{equation}
where $\lambda$ and $\mu$ are Lagrangian multipliers and $\langle \cdot \rangle = \frac{1}{2\pi} \int_{0}^{2\pi} ( \cdot ) d\theta$. The first term of the cost function $-\Gamma^\prime(\psi_*)$ is the speed of synchronization, the second term constrains the actuation energy and the third term enforces the synchronization condition.

Once the phase sensitivity is obtained, this optimization problem can be solved analytically using the calculus of variations \citep{zlotnik2013optimal}. Here, without loss of generality, we consider the fixed point, $\psi_*=0$ since the long-term synchronization dynamics is independent of the choice of initial phase. The phase-sensitivity-based actuation waveform is obtained analytically as
\begin{equation}
     \eta_{Z}(\theta;\Delta \omega/\epsilon) = -\frac{Z^\prime(\theta)}{2\lambda} - \frac{(\Delta \omega /\epsilon) Z(\theta)}{\langle Z^2 \rangle},\quad \lambda = \frac{1}{2}\sqrt{\frac{\langle(Z^\prime)^2\rangle}{1-\frac{(\Delta \omega/\epsilon)^2}{ \langle Z^2 \rangle}}},
    \label{eq:optimal_waveform}
\end{equation}
where $Z^\prime (\theta)$ is the derivative of the effective phase sensitivity function with respect to $\theta$. Using equation~\ref{eq:optimal_waveform}, we can determine the actuation waveform that can quickly synchronize to a frequency $\omega_f$ and the associated magnitude of perturbation, $\epsilon$. While this phase-sensitivity-based waveform is optimized for fast frequency modification, it is to be noted that this is theoretically valid within the linear regime in the vicinity of a perfect limit cycle. Since we are concerned with turbulent oscillatory flows, we investigate the efficacy of this phase-sensitivity-based actuation waveform by performing numerical simulations.
	
\section{Controlled cavity flows}
\label{sec:results}

We perform active flow control of cavity flows to suppress the pressure fluctuations using unsteady blowing at the leading edge of the cavity by modifying the dominant vortex convection frequency of $St_L=1.32$ using a different actuation frequency close to the dominant frequency. To this end, we consider actuation frequencies that are within $20\%$ of $St_L=1.32$ as $St_L=1.06,1.19,1.45$ and $1.58$. We consider two actuation waveforms for unsteady flow control: a sinusoidal waveform at the selected actuation frequency and the phase-sensitivity-based actuation waveform for fast frequency modification in equation~\ref{eq:optimal_waveform} at the same fundamental frequency. In addition to actuation frequency, spanwise actuation wavenumber is another important parameter for flow control. Further, a previous control study for the same flow conditions by \cite{liu2021unsteady} used resolvent analysis to identify the spanwise actuation wavenumber responsible for efficient fluctuation suppression. For actuation frequencies around $St_L=1.32$, they identified that the effective control of fluctuations corresponds to the spanwise actuation wavenumbers of $\beta=\pi$ and $2\pi$, respectively. Hence, for the present analysis, we compute the phase response and phase-based control and compare the performance with that using a sinusoidal waveform using the spanwise wavenumbers of $\beta=\pi$ and $2\pi$.

\begin{figure}
    \centering
    \includegraphics[width=\textwidth]{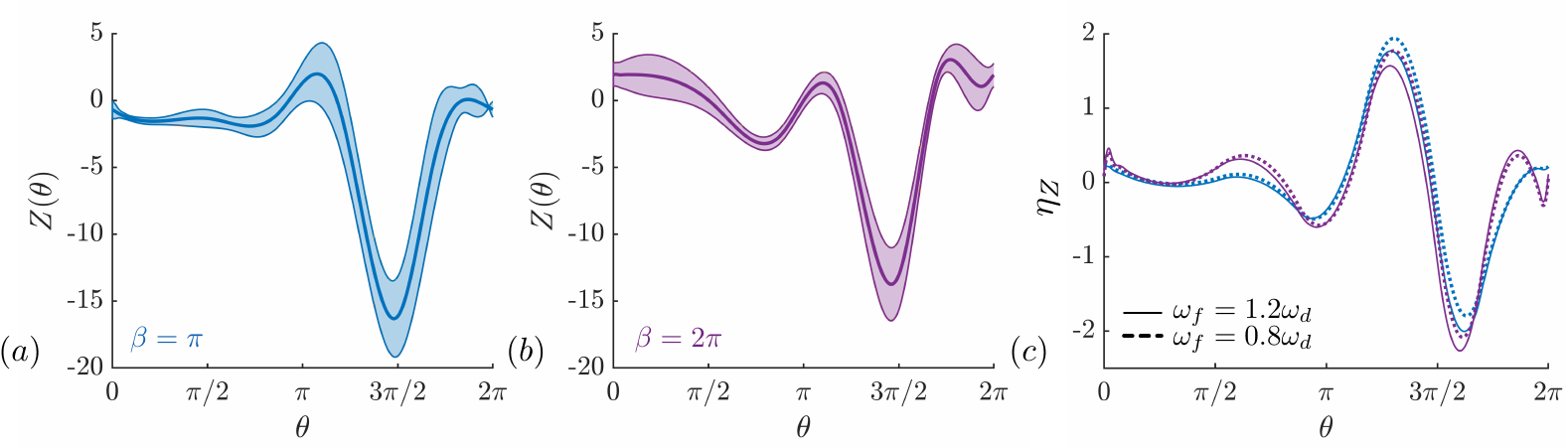}
    \caption{(a)-(b) Phase sensitivity functions computed using the pulse-train method for $\beta=\pi$ and $2\pi$, respectively. (c) Phase-based actuation waveform for fast synchronization for actuation frequencies, $\omega_f=0.8\omega_d$ and $1.2\omega_d$ for spanwise wavenumbers of $\beta=\pi,2\pi$ for a momentum coefficient of $C_\mu=0.02$. }
    \label{fig:Ztheta}
\end{figure}

\subsection{Phase sensitivity functions}

We assess the first-order phase response of the cavity flows about the dominant oscillation cycle corresponding to the convection of the large-scale spanwise structures in the cavity flow using the pulse-train method discussed in section~4.2. We introduced impulsive perturbations from the actuation at the leading edge with spanwise actuation wavenumbers $\beta=\pi$ and $2\pi$ for each phase at every time the trajectory reaches $\theta=\theta_0$ defined based on the spanwise DMD modes to compute the averaged phase response for each phase $\theta_0$. We considered such simulations for various $\theta_0$ for both the spanwise wavenumber. 

The phase sensitivity functions obtained after perturbing for 13 cycles are shown in figure~\ref{fig:Ztheta}(a)-(b). The trend of the phase-sensitivity function is confirmed to converge after 8 perturbation cycles. The plotted phase sensitivity functions are computed using the average of the last five impulse perturbations, and the shaded area represents the standard deviation of the last five cycles of impulse perturbation. We note that since the phase sensitivity is computed based on the spanwise-averaged flowfield, the phase sensitivity based on three-dimensional perturbation characterizes the most sensitive phase where the three-dimensional perturbation can cause delay/advancement of the two-dimensional large-scale spanwise vortex convection. We observe that for both $\beta=\pi$ and $2\pi$, the phase sensitivity function shows a similar trend. The magnitude of phase sensitivity is also similar, albeit slightly higher at the sensitive phase for $\beta=\pi$. For both the wavenumbers, the most sensitive phase is indicated as $\theta=3\pi/2$, where the impulse response is observed to cause phase delay. We note that while the absolute value of $\theta$ is dependent on the chosen phase space, the flow state associated with a phase and the waveform of the phase sensitivity function encodes the effect of the perturbation on the respective flow state. From figure~\ref{fig:phase_dynamics}((d),(h)), $\theta=3\pi/2$ is mapped to the flow state when the large-scale spanwise vortex is convected past the mid-cavity. This phase precedes the phase, $\theta=0$, where the strong spanwise vortex is observed at $x/D=2$. This indicates that introducing a small three-dimensional impulse perturbation at both $\beta=\pi$ and $2\pi$ at the leading edge at $\theta=3\pi/2$ is convected through the shear layer and significantly modifies the formation of the strong spanwise vortex and causes a delay of the vortex-convection process. 

Using the phase sensitivity functions, we then compute the actuation waveform $(\eta_Z)$ for fast frequency modification using equation~\ref{eq:optimal_waveform} for actuation frequencies within $20\%$ range of the dominant frequency for the magnitude $\epsilon$ corresponding to $C_\mu=0.02$. The waveforms for both $\beta=\pi$ and $2\pi$ for $\omega_f=0.8\omega_d$ and $1.2\omega_d$ are shown in figure~\ref{fig:Ztheta}(c). While we show the waveforms for the two frequencies, the waveforms associated with $\omega_f=0.9\omega_d$ and $1.1\omega_d$ lie between them. Since the phase sensitivity functions are similar, the fast modification waveforms are also similar for $\beta=\pi$ and $2\pi$. These waveforms resemble a stretched pulse with a significant amount of actuation concentrated between $\theta=\pi$ and $2\pi$, targeting the sensitive phase of $\theta=3\pi/2$ with minimal actuation input for the other phases. 

\begin{figure}
\centering
    \rotatebox{90}{
        \begin{minipage}{\textheight}
            \centering
            \includegraphics[width=0.95\textheight]{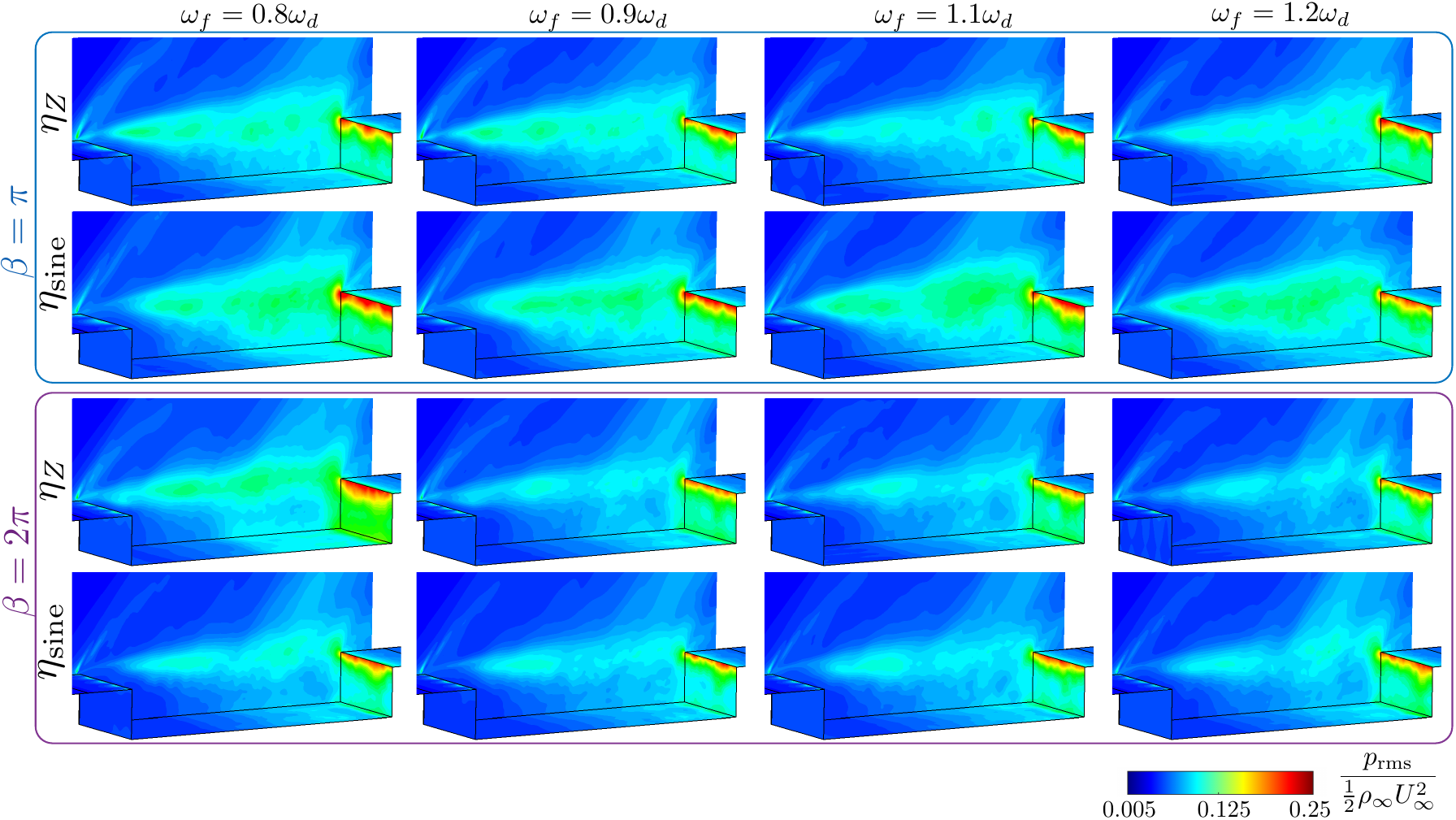}
           \caption{Distribution of $p_{\textrm{rms}}$ along the cavity walls and a spanwise slice of $z=0$ for controlled flows using phase-sensitivity-based and sinusoidal actuation waveforms at $C_\mu=0.02$ for actuation frequencies $\omega_f=0.8\omega_d,0.9\omega_d,1.1\omega_d$ and $1.2\omega_d$ and spanwise wavenumbers $\beta=\pi$ and $2\pi$.}
             \label{fig:prms_control}
        \end{minipage}
        }
\end{figure}

\subsection{Open-loop flow control}

We can now perform open-loop flow control at the actuation frequencies of $\omega_f=0.8\omega_d,0.9\omega_d,1.1\omega_d$ and $1.2\omega_d$ at $\beta=\pi$ and $2\pi$ using both $\eta_Z$ and a sinusoidal waveform, $\eta_{\textrm{sine}}$ at the same actuation frequency by introducing unsteady blowing at the leading edge actuator. 

The distribution of r.m.s. fluctuations on the cavity walls and a spanwise slice at plane $z=0$ for all the cases of actuation frequencies using both $\beta=\pi$ and $2\pi$ and waveforms, $\eta_Z$ and $\eta_{\textrm{sine}}$ are shown in figure~\ref{fig:prms_control}. Overall, we observe a significant reduction in the pressure fluctuations for the considered controlled cases compared to the uncontrolled flowfield (seen in figure~\ref{fig:baseflow}(b)). In contrast to the uncontrolled flowfield wherein large fluctuations are observed on the aft-wall and the shear layer region, the controlled flows show a reduction in $p_{\textrm{rms}}$ along both the shear layer and the cavity walls. 

Let us consider the controlled flow scenarios with spanwise actuation wavenumber of $\beta=\pi$ and compare the effect of actuation waveform on the $p_{\textrm{rms}}$ distribution. The main difference is observed within the cavity along the shear layer for all the actuation frequencies. We observe lower r.m.s. fluctuations along the shear layer across the cavity length when actuated using phase-sensitivity-based waveform $\eta_Z$. The difference between the control performance of the waveforms is more significant for higher actuation frequencies, $\omega_f=1.1\omega_d$ and $1.2\omega_d$. The difference can be inferred from the narrower transverse spatial extent of large fluctuations over the shear layer when actuated using $\eta_Z$, whereas the actuation using sinusoidal waveform shows a broader transverse spatial extent of large fluctuations. This narrower spatial region at higher actuation frequencies could indicate the weakening of the large-scale vortical structures resulting in reduction of pressure fluctuations. Overall, for actuation using $\beta=\pi$, the phase-sensitivity-based actuation waveform achieves more reduction of $p_{\textrm{rms}}$ with actuation frequencies higher than the dominant frequency of the uncontrolled flow. 

\begin{figure}
    \centering
    \includegraphics[width=0.95\textwidth]{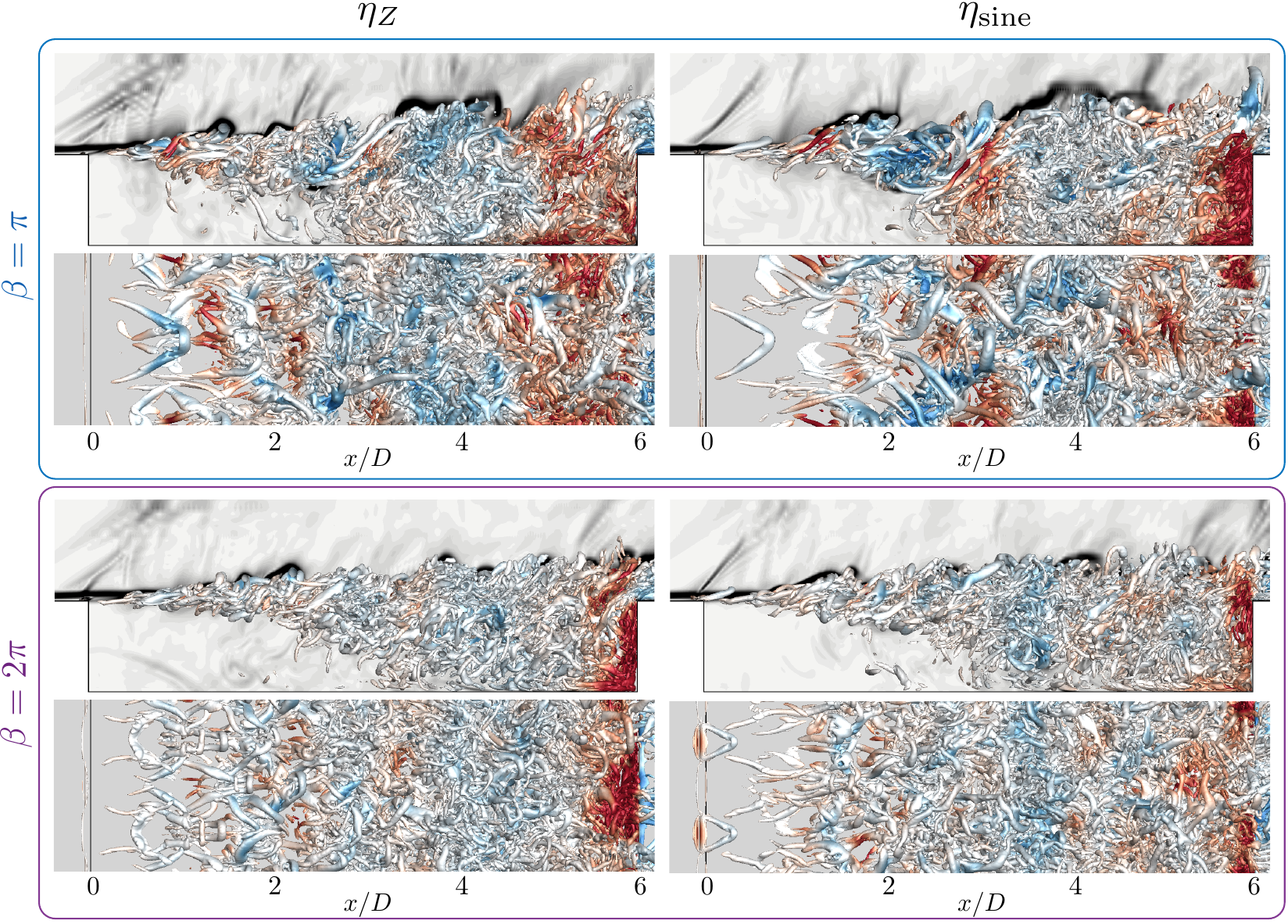}
    \caption{Instantaneous $xy-$ and $xz-$ views of controlled flowfields visualized using isosurface of $Q$-criterion $Q=14$ colored using the contour of pressure coefficient with numerical schlieren of spanwise slice $z=0$ in the background. Control using phase-sensitivity-based $\eta_Z$ and sinusoidal waveforms $\eta_{\textrm{sine}}$ using (top) $\beta=\pi$ and $\omega_f=1.1\omega_d$ and using (bottom) $\beta=2\pi$ and $\omega_f=1.2\omega_d$.}
    \label{fig:controlled_flow}
\end{figure}

When the spanwise actuation wavenumber is $\beta=2\pi$, the $p_{\textrm{rms}}$ distribution of the controlled flows show similar distribution for both the actuation waveforms, $\eta_Z$ and $\eta_{\textrm{sine}}$ when actuated using $\omega_f=0.9\omega_d,1.1\omega_d$ and $1.2\omega_d$. In contrast to $\omega_f=0.8\omega_d$, the $p_{\textrm{rms}}$ distribution shows larger fluctuations over the shear layer and the cavity walls when actuated using phase-sensitivity-based waveform $\eta_Z$ compared to the sinusoidal actuation waveform $\eta_{\textrm{sine}}$. Even when using sinusoidal actuation waveform, $\omega_f=0.8\omega_d$ shows larger fluctuations over the shear layer than the other actuation frequencies. Additionally, high-frequency actuation $\omega_f=1.2\omega_d$ shows a narrower transverse extent of large fluctuations when compared with lower actuation frequencies along the shear layer, especially till the mid-cavity from the leading edge. Further, when comparing the control performance of the two actuation wavenumbers $\beta=\pi$ and $2\pi$, we observe that $\beta=2\pi$ achieves more attenuation of pressure fluctuations within the cavity when actuated using both the waveforms. 

To further investigate the flow physics of the controlled flows, we visualize the instantaneous flowfields in figure~\ref{fig:controlled_flow} for a couple of actuator parameter combinations using both waveforms: (top row) $\beta=\pi$ and $\omega_f=1.1\omega_d$, (bottom row) $\beta=2\pi$ and $\omega_f=1.2\omega_d$. When compared with the uncontrolled flowfields shown in figure~\ref{fig:baseflow}(a) and figures~\ref{fig:phase_dynamics}(a)-(h), the controlled flows exhibit weakened the large spanwise structures. We observe the presence of more small-scale structures close to the leading edge and shear-layer thickening due to the introduction of leading-edge actuation. For all the shown controlled flows, we observe a suppression of fluctuations in the shear layer, which is also seen as weaker compression waves in the background schlieren than the uncontrolled flow. The wall-normal leading-edge actuation introduces streamwise vortices into the shear layer as seen from the $xz$-view of the controlled flowfields.

\begin{figure}
    \centering
    \includegraphics[width=0.95\textwidth]{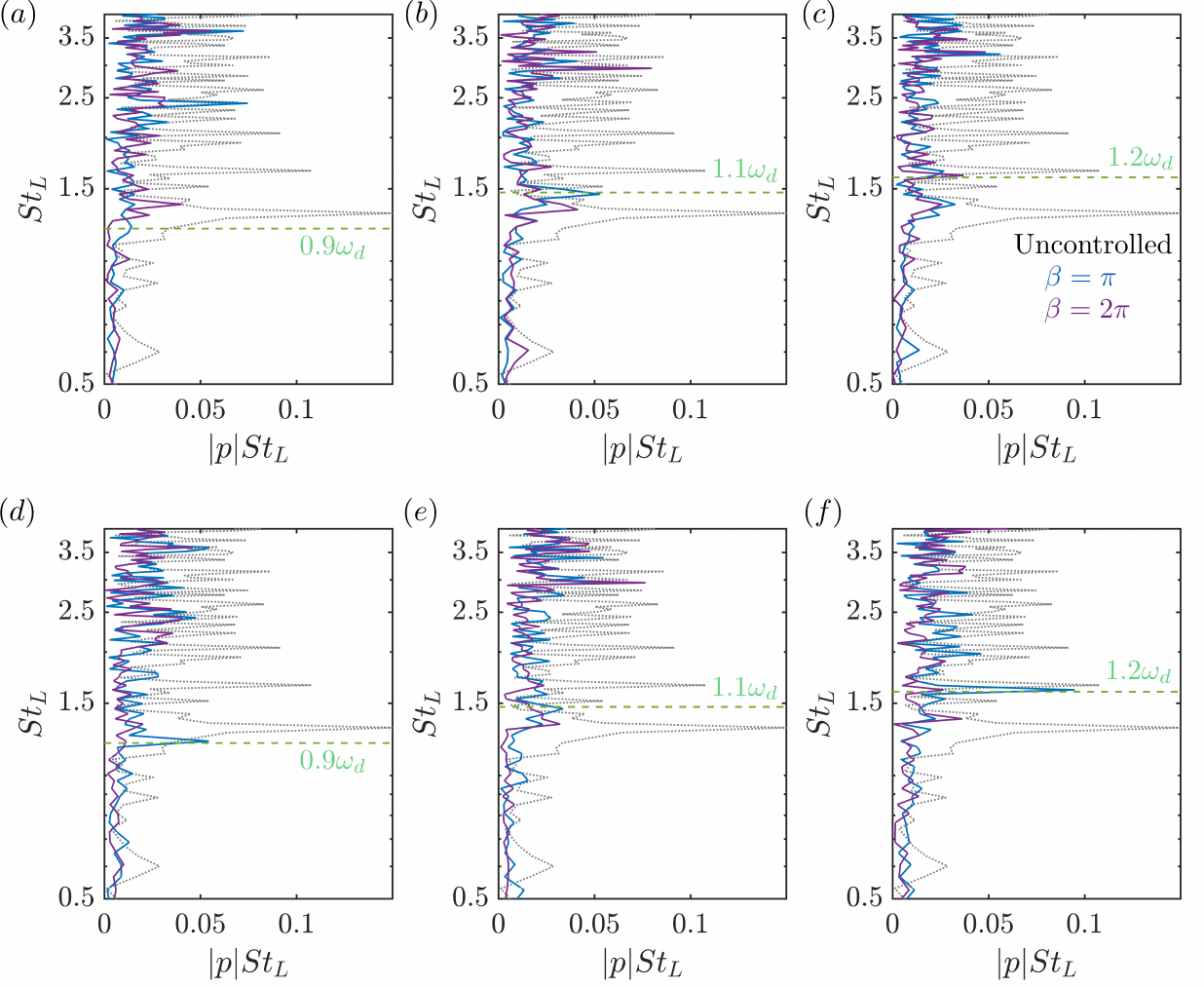}
    \caption{Premultiplied pressure spectra at a probe placed mid-cavity along the shear layer $(x,y,z)/D=(3,0,1)$ for uncontrolled flow shown in black dotted line and controlled flows using $\beta=\pi$ (blue) and $2\pi$ (purple). Controlled cases correspond to the actuation frequencies, $0.9\omega_d,1.1\omega_d$ and $1.2\omega_d$ (shown as green dashed lines) actuated using $\eta_Z$ (a)-(c) and $\eta_{\textrm{sine}}$ waveforms (d)-(f), respectively.}
    \label{fig:fft_control}
\end{figure}

\begin{figure}
    \centering
    \includegraphics[width=\textwidth]{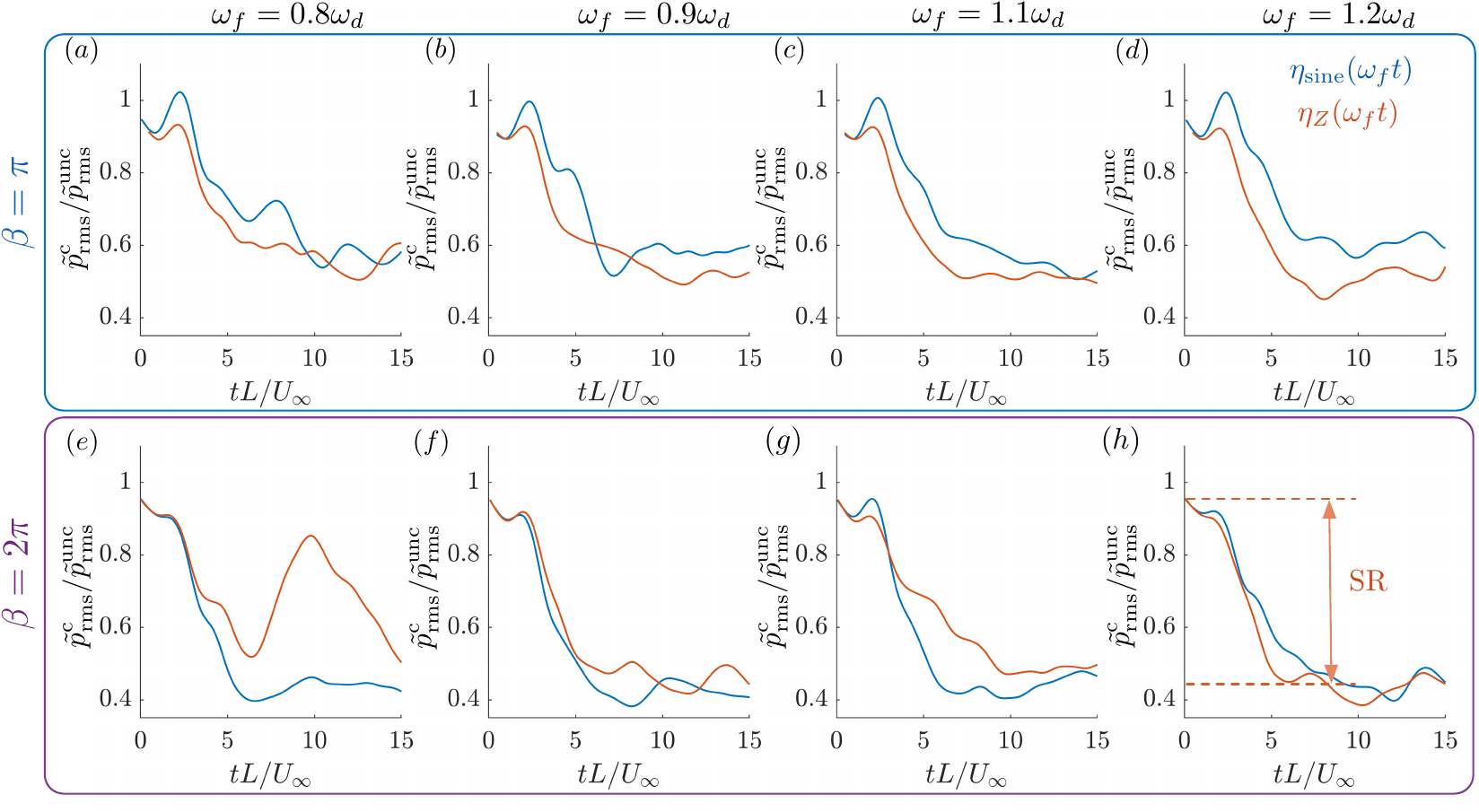}
    \caption{Temporal variation of integrated $p_{\textrm{rms}}$ ratio between the controlled and uncontrolled flows when actuated using spanwise wavenumbers of $\beta=\pi$ (a)-(d) and $\beta=2\pi$ (e)-(h) and actuation frequencies of $\omega_f=0.8\omega_d,0.9\omega_d,1.1\omega_d$ and $1.2\omega_d$, respectively for both the actuation waveforms, $\eta_Z$ and $\eta_{\textrm{sine}}$.}
    \label{fig:speed}
\end{figure}
 
\begin{figure}
    \centering
    \includegraphics[width=0.7\textwidth]{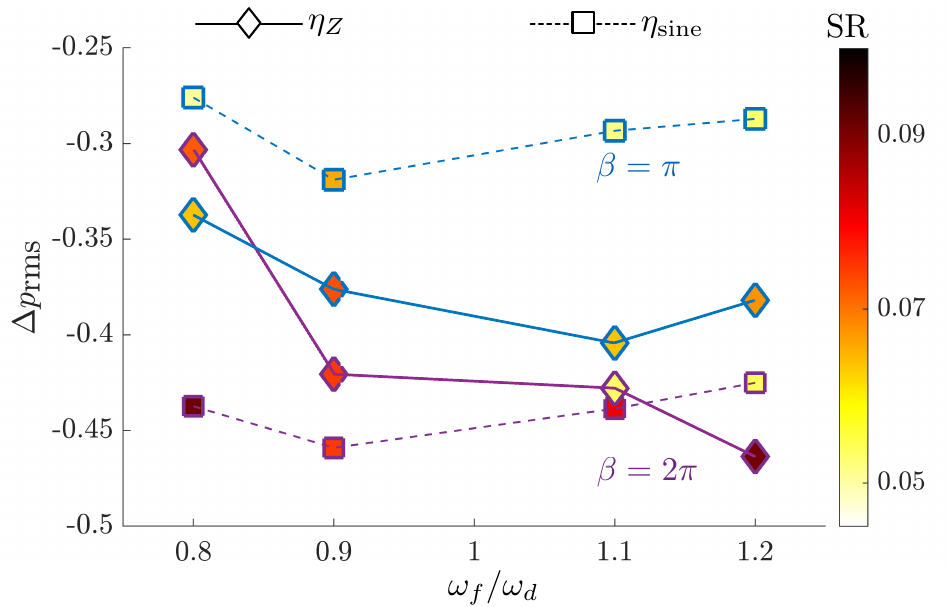}
    \caption{\textrm{Suppression of pressure fluctuations ($\Delta p_{\textrm{rms}}$) and the corresponding speed of reduction for the considered control scenarios (SR)}}
    \label{fig:control_performance}
\end{figure}

When actuated with $\beta=\pi$ (top row), the controlled flows using both the waveforms show stronger vortices in the shear layer seen from the envelope of compression waves ($xy$-view) when compared with cavity flows actuated with $\beta=2\pi$ (bottom row). This was observed as a wider transverse region with large fluctuations in the $p_{\textrm{rms}}$ distribution in figure~\ref{fig:prms_control}. For $\beta=\pi$ case, actuation using $\eta_Z$ results in the emergence of a larger number of small-scale structures till $x/D=2.5$ when compared with $\eta_{\textrm{sine}}$ actuation (see $xz$-view). When actuated using sinusoidal waveform at the same actuation frequency of $\omega_f=1.1\omega_d$, we observe longer streamwise elongated structures about $x/D=2$ whereas we observe shorter streamwise vortices and their mixing around $x/D=2$ resulting in a more stable shear layer when actuated using $\eta_Z$. Thus, this also weakens the effect of impingement of these vortices on the aft-wall as can be observed from the lower $p_{\textrm{rms}}$ fluctuations near the aft-wall in figure~\ref{fig:prms_control} for $\beta=\pi$. 

 Let us consider the controlled flow scenario with $\beta=2\pi$. The pressure distributions show similar variation for $\omega_f=0.9\omega_d,1.1\omega_d$ and $1.2\omega_d$. Here, we visualize the controlled flow for $\omega_f=1.2\omega_d$ for $\beta=2\pi$ as a representative case. Since the actuation is at a higher spanwise wavenumber and frequency, we observe more number of smaller streamwise vortices for $\beta=2\pi$ (see $xz$-view for $\beta=2\pi$ in figure~\ref{fig:controlled_flow}). Similar to the actuation cases with $\beta=\pi$, we observe a larger number of small-scale structures and the interaction among them from $x/D=0$ to $2$ when actuated using $\eta_Z$ waveform when compared with $\eta_{\textrm{sine}}$ waveform. Both sinusoidal and phase-sensitivity-based waveforms, when actuated at $\beta=2\pi$, have a more stabilizing effect on the shear layer, resulting in small-scale weaker vortical structures, thereby attenuating pressure fluctuations within the cavity.

 Since the motivation of the control strategy is to weaken the dominant frequency by introducing a slightly different time scale, we visualize the premultiplied pressure spectra at a mid-cavity probe placed along the shear layer for a variety of actuation cases in figure~\ref{fig:fft_control}. The pressure spectra are computed using 40 oscillation cycles with respect to $St_L=1.32$. All the controlled cases effectively suppress the dominant tone of $St_L=1.32$ and the overall pressure fluctuations. The actuation frequency for each control scenario is shown using the green dashed line. A secondary peak with much lower energy is observed at $St_L=1.32$, indicating a successful frequency detuning. For the same $\beta=\pi$, actuation using $\eta_Z$ results in a secondary peak with lower energy content at $St_L=1.32$ similar to the case of  $\eta_Z$. However, we also observe a weaker peak in the spectra at the actuation frequency when compared with the sinusoidal actuation waveform, thereby overall reduction in fluctuations in the cavity when actuated with phase-sensitivity-based waveform at $\beta=\pi$. This can also be attributed to the presence of the smaller streamwise vortices in the shear layer, which results in the breakdown of the large-scale spanwise vortices being more effective than those introduced with sinusoidal actuation, weakening the feedback loop. Like $\beta=\pi$, actuation at $\beta=2\pi$ can effectively suppress the dominant tone at $St_L=1.32$. Further, a more apparent reduction in overall fluctuations when compared with $\beta=\pi$ is seen for sinusoidal waveform over a frequency band of $1.5<St_L<2.5$. Thus, introducing actuation at a frequency slightly different from the dominant frequency can suppress the overall fluctuations by disrupting the feedback loop. 

The phase-sensitivity-based actuation waveform is designed to quickly modify the flow frequency. While the previous analysis so far considered the asymptotic statistics of controlled flows, we now investigate how quickly we can suppress the pressure fluctuations within the cavity. We visualize the ratio of the integrated $p_{\textrm{rms}}$ between the controlled and uncontrolled flows as, $\tilde{p}^{\textrm{c}}_{\textrm{rms}}/\tilde{p}^{\textrm{unc}}_{\textrm{rms}}$ within the cavity using the domain of integration as $(x,\,y,\,z)/D \in ([0,6],[-1,1.5],[0,2])$ for all the actuation scenarios. The speed of reduction (SR) is quantified using 
\begin{equation}
    \textrm{SR} = \frac{\frac{\tilde{p}^{\textrm{c}}_{\textrm{rms}}}{\tilde{p}^{\textrm{unc}}_{\textrm{rms}}}\bigg|_{t=0}-\frac{\tilde{p}^{\textrm{c}}_{\textrm{rms}}}{\tilde{p}^{\textrm{unc}}_{\textrm{rms}}}\bigg|_{t=t_0}}{t_0},
\end{equation}
where $t_0$ is the time instant of first inflection point for $p^{\textrm{c}}_{\textrm{rms}}/p^{\textrm{unc}}_{\textrm{rms}}$. An instance for computing SR is shown in figure~\ref{fig:speed}(h).

\begin{table}[h]
  \begin{center}
\def~{\hphantom{0}}
  \begin{tabular}{l @{\hspace{70pt}} c @{\hspace{30pt}} c @{\hspace{30pt}} c@{\hspace{30pt}} c @{\hspace{30pt}} c}
      \hline
      \noalign{\vskip 3pt}
      Cases & $\beta$ & $St_c$ & $\tilde{p}_\text{rms}$ & $\Delta \tilde{p}_\text{rms}\times 100\%$ & SR \\
      [3pt]
      \hline
      \noalign{\vskip 3pt}
      Uncontrolled & - & - & 2.41 & - & - \\
      [3pt]
      \hline
      \noalign{\vskip 3pt}
      \multirow{8}{*}{$\eta_Z$} & $\pi$ & 1.06 & 1.60 & -33.7\% & 0.064 \\
      & $\pi$ & 1.19 & 1.51 & -37.6\% & 0.073 \\
      & $\pi$ & 1.45 & 1.44 & -40.5\% & 0.064 \\
      & $\pi$ & 1.58 & 1.49 & -38.2\% & 0.068 \\
      [5pt]
      & $2\pi$ & 1.06 & 1.68 & -30.3\% & 0.072 \\
      & $2\pi$ & 1.19 & 1.40 & -42.1\% & 0.075 \\
      & $2\pi$ & 1.45 & 1.38 & -42.8\% & 0.053 \\
      & $2\pi$ & 1.58 & 1.29 & -46.4\% & 0.091 \\
      \noalign{\vskip 3pt}
      \hline
      \noalign{\vskip 3pt}
      \multirow{8}{*}{$\eta_{\textrm{sine}}$} & $\pi$ & 1.06 & 1.75 & -27.4\% & 0.050 \\
      & $\pi$ & 1.19 & 1.64 & -31.9\% & 0.066 \\
      & $\pi$ & 1.45 & 1.71 & -29.3\% & 0.051 \\
      & $\pi$ & 1.58 & 1.72 & -28.7\% & 0.052 \\
      [5pt]
      & $2\pi$ & 1.06 & 1.36 & -43.8\% & 0.091 \\
      & $2\pi$ & 1.19 & 1.30 & -45.9\% & 0.075 \\
      & $2\pi$ & 1.45 & 1.35 & -43.9\% & 0.079 \\
      & $2\pi$ & 1.58 & 1.39 & -42.5\% & 0.091 \\
      \noalign{\vskip 3pt}
      \hline
    \end{tabular}
  \end{center}
  \caption{Summary of unsteady cavity flow control scenarios}
  \label{tab:kd}
\end{table}

When using the actuation wavenumber $\beta=\pi$ ((a)-(d)), the phase-sensitivity-based actuation waveform achieves quicker suppression of pressure fluctuations compared to that of sinusoidal actuation, as predicted by the phase-based analysis. Further, the control performance is better at higher frequencies (c)-(d) than the dominant frequency, as the $p_{\textrm{rms}}$ ratio achieves a lower value much quicker than at lower actuation frequencies. For these high-frequency actuation scenarios, the inflection point is observed at about six convective time units based on the cavity length for the phase-sensitivity-based waveform. When $\beta=2\pi$, for $\omega_f=0.9\omega_d,1.1\omega_d$ and $1.2\omega_d$ the initial variation of the integrated $p_{\textrm{rms}}$ ratio is similar for both phase-sensitivity-based and sinusoidal waveforms, and we observe similar reduction and inflection points for $\omega_f=0.9,1.2\omega_d$. However, for $\omega_f=1.1\omega_d$, we observe that the sinusoidal waveform achieves quicker suppression than the phase-sensitivity-based waveform. In addition, $\omega_f=0.8\omega_d$ shows an oscillating $p_{\textrm{rms}}$ ratio for phase-sensitivity-based waveform, suggesting that the actuation energy might not be enough to achieve frequency modification to a lower frequency for $\beta=2\pi$.

\begin{figure}[h]
    \centering
    \includegraphics[width=0.9\textwidth]{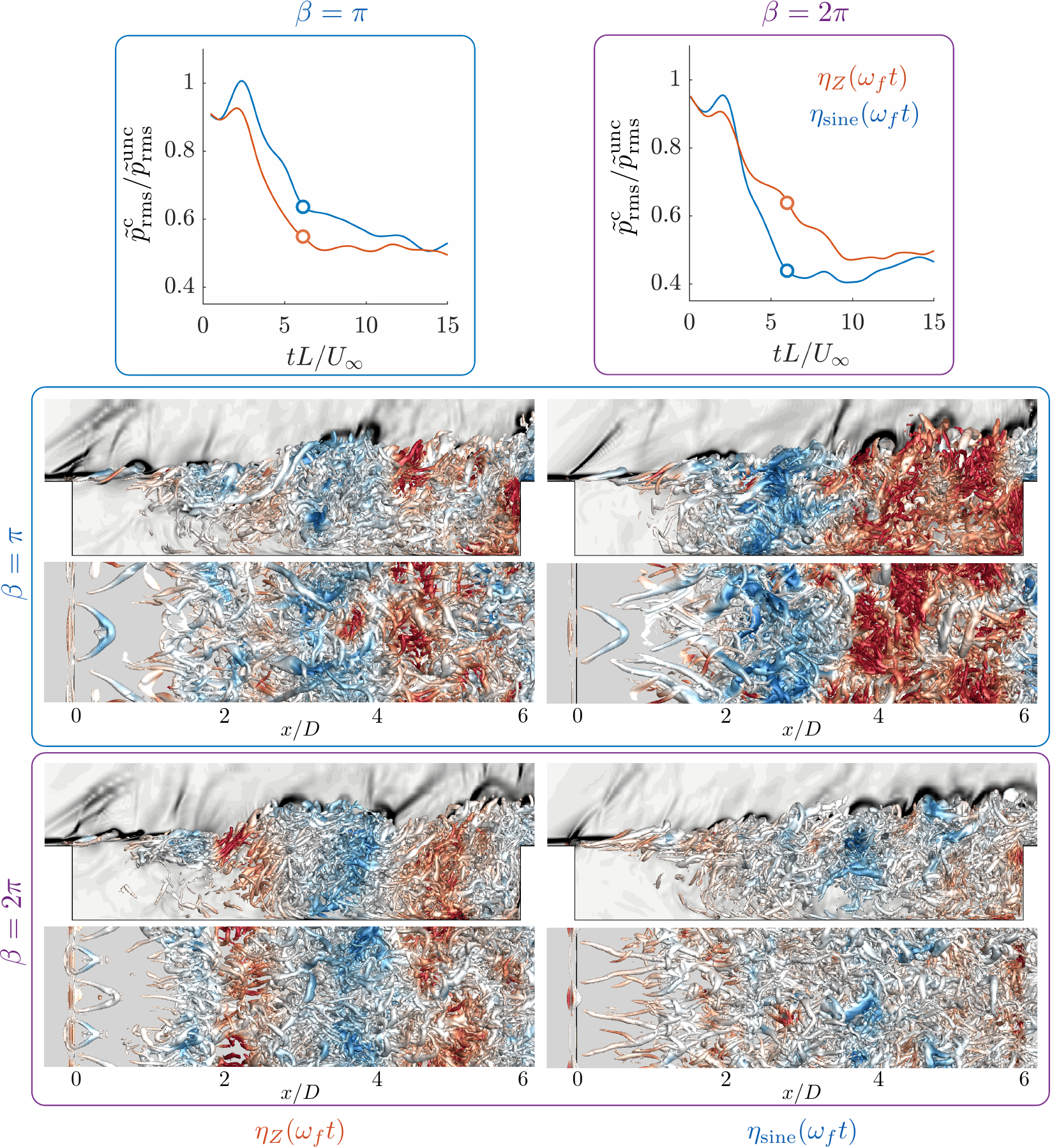}
    \caption{Instantaneous $xy-$ and $xz-$ views of controlled flowfields visualized using isosurface of $Q$-criterion $(Q=14)$ colored using the contour of pressure coefficient with numerical schlieren of spanwise slice $z=0$ in the background at $tL/U_{\infty}=6$. Control using phase-sensitivity-based $\eta_Z$ and sinusoidal waveforms $\eta_{\textrm{sine}}$ using (top) $(\beta,\omega_f)=(\pi,1.1\omega_d)$ and (bottom) $(2\pi,1.1\omega_d)$.}
    \label{fig:example}
\end{figure}

The summary of control performance combining the overall reduction of pressure fluctuations and the speed of reduction for all the considered actuation scenarios is provided in table~\ref{tab:kd} and is visualized in figure~\ref{fig:control_performance}. The data points in figure~\ref{fig:control_performance} are colored using the corresponding speed of reduction. We observe that phase-sensitivity-based actuation waveform outperforms sinusoidal waveform both in terms of the magnitude of reduction and the speed of reduction for $\beta=\pi$. The maximum reduction observed for phase-sensitivity-based waveform is about 41\% for $(\beta,\omega_f)=(\pi,1.1\omega_d)$ and a similar reduction of 38\% for $(\beta,\omega_f)=(\pi,1.2\omega_d)$. The controlled case with sinusoidal waveform achievers a maximum reduction of 32\% for $(\beta,\omega_f)=(\pi,0.9\omega_d)$ and other similar reductions of 29\% for $(\beta,\omega_f)=(\pi,1.1\omega_d)$ and $(\pi,1.2\omega_d)$. Thus, phase-sensitivity-based actuation waveform achieves 10\% higher attenuation of pressure fluctuations compared with sinusoidal waveform at $\beta=\pi$. We observe an interesting trend for $\beta=2\pi$. The sinusoidal actuation waveform achieves a maximum reduction of 46\% for $(\beta,\omega_f)=(2\pi,0.9\omega_d)$ followed by similar reductions of 44\%, 44\% and 43\% for $(\beta,\omega_f)=(2\pi,0.8\omega_d),(2\pi,1.1\omega_d),(2\pi,1.2\omega_d)$, respectively. The phase-sensitivity-based actuation waveform suppresses the pressure fluctuations by 46\% for $(\beta,\omega_f)=(2\pi,1.2\omega_d)$ and similar reductions of 42\% and 43\% for $(\beta,\omega_f)=(2\pi,0.9\omega_d),(2\pi,1.1\omega_d)$, respectively. However, for $(\beta,\omega_f)=(2\pi,0.8\omega_d)$, phase-sensitivity-based waveform suppresses the pressure fluctuations by 30\%. This indicates that an overall higher frequency actuation achieves better reduction when using a phase-sensitivity-based actuation waveform. Further, for $\beta=2\pi$, both the magnitude and speed of pressure fluctuation reduction are similar for both sinusoidal and phase-sensitivity-based actuation waveforms for $(\beta,\omega_f)=(2\pi,0.9\omega_d)$ and $(2\pi,1.2\omega_d)$. However, for $(\beta,\omega_f)=(2\pi,1.1\omega_d)$, sinusoidal waveform achieves a quicker reduction of pressure fluctuations within the cavity. 

Let us investigate the flow physics responsible for quick attenuation of the pressure fluctuations within the cavity based on the instantaneous flowfields in figure~\ref{fig:example}. The flowfields of the actuation scenarios correspond to $tL/U_{\infty}=6$, where the actuation is turned on at $tL/U_{\infty}=0$. We consider $\omega_f=1.1\omega_d$ as we observe a different trend of SR for phase-sensitivity-based and sinusoidal waveforms for $\beta=2\pi$. Since the control performance is similar at $\beta=\pi$ for both $\omega_f=1.1\omega_d$ and $1.2\omega_d$, the same observations drawn also hold for $\omega_f=1.2\omega_d$. The initial condition for introducing actuation is the same for all the control scenarios. For $\beta=\pi$, from the instantaneous flowfields, the phase-sensitivity-based actuation waveform achieves a breakdown of the large-scale spanwise vortex structure within just six convective time units, whereas the sinusoidal waveform shows a more gradual change in weakening the spanwise vortex structure near $x/D=3$. Actuation using $\eta_Z$ shows a more compact and larger number of streamwise vortices at $1<x/D<2$ when compared with $\eta_{\textrm{sine}}$ actuation, which comprises of elongated and fewer streamwise vortices. For $\eta_Z$, these smaller streamwise vortices convect along the shear layer and quickly modify the spanwise vortex roll-up through spanwise mixing at $x/D>1.75$, preceding the streamwise location of the large-scale spanwise vortex as identified by the phase sensitivity function in figure~\ref{fig:phase_dynamics}. Since the sinusoidal actuation introduced elongated streamwise vortices, we observe the emergence of small-scale structures and a gradual loss of coherence and mixing past $x/D>2$.

For $\beta=2\pi$, as shown in the $xz$-view, the phase-sensitivity-based waveform actuation similarly introduces a more compact and larger number of streamwise vortices at $0<x/D<1$ when compared with sinusoidal actuation waveform. While the sinusoidal actuation waveform breaks the large-scale spanwise vortices, we observe strong fluctuations and the emergence of coherence for phase-sensitivity-based actuation waveform at streamwise location $1.5<x/D<2.5$. The streamwise vortices are observed to align in a closely packed manner in the spanwise direction at $x/D=1.5$. These closely packed vortices appear to cancel out their influence, reducing the actuation effect. This is also seen as a local thickening of the shear layer at $x/D=1.5$ in the $xy$-view of the flowfield. Hence, for a few specific $(\beta,\omega_d)=(2\pi,1.1\omega_d)$, it is possible that sinusoidal actuation achieves quicker suppression than phase-sensitivity-based actuation waveform. However, we observe similar control performance for higher actuation frequency of $1.2\omega_d$.

\subsection{Outlook}

The current phase-based analysis is based on spanwise-averaged flow physics and can capture/modify the large-scale spanwise vortex convection dynamics within the cavity. The phase-based control objective is framed to modify this dominant frequency, disrupting the feedback loop and, hence, indirectly reducing the pressure fluctuations within the cavity. For the current cavity flow control, both the sinusoidal and phase-sensitivity-based actuation waveform achieve a similar reduction of $46\%$ with an actuation wavenumber of $\beta=2\pi$. While the phase-based control waveform does not further enhance the reduction speed, we attributed this to the three-dimensional alignment and the nonlinear interaction of actuation at this wavenumber. Further, this observation is specific to the current problem setup, and the effect of variation of cavity geometry, inflow conditions, and leading-edge boundary layer thickness on the phase sensitivity and phase-based control is yet to be investigated. In addition, the performance of the phase-sensitivity-based actuation waveform is significantly better than the sinusoidal waveform for $\beta=\pi$. A combination of actuation wavenumbers and frequencies is yet to be investigated as a possibility to achieve an even higher attenuation of pressure fluctuations. 

Phase-based analysis for turbulent flows can serve as a guideline for formulating timing-based flow control strategies for complex, unsteady flows. Exploring connections between the resolvent analysis and phase-based analysis could be advantageous for designing efficient flow control strategies as resolvent analysis identifies the most amplified response to perturbations while phase-based analysis identifies the most sensitive time to introduce actuation. \cite{liu2021unsteady} identified that $(\beta=2\pi,St_L=2.49)$ achieves a pressure fluctuation reduction of 52\% on the cavity walls guided by resolvent analysis. There also have been previous studies to identify secondary actuator location on the bottom cavity walls supplementing the leading edge actuation \cite{godavarthi2024windowed}. Thus, a combination of resolvent-based actuation at the leading edge and phase-based optimal control at a secondary actuation location could achieve a \textit{quicker} and higher attenuation of pressure fluctuations within the cavity and is to be studied.

Further, with recent developments in data-driven techniques, machine learning has been used to identify and model the dynamics in low-order coordinates \citep{lusch2018deep,champion2019data,linot2020deep}. Recently, phase-based analysis has been used with machine learning methods to obtain phase response and the description of periodic and transient flows \citep{smith2024cyclic,fukami2024data,yawata2024phase}. With these promising advancements, phase-based analysis can potentially develop unsteady flow control strategies for unsteady flows.

\section{Conclusions}
\label{sec:conclusions}

We considered open-loop flow control to attenuate pressure fluctuations in spanwise-periodic supersonic turbulent cavity flow with an incoming Mach number of 1.4 and Reynolds number of 10000. We aimed to achieve the suppression of fluctuation by modifying the dominant frequency of the large-scale vortex convection in the cavity. We extended the phase-reduction approach to turbulent flows to identify the phase response about the dominant frequency in terms of delay/advancement of the vortex convection cycle and designed a phase-based actuation waveform to quickly modify the dominant frequency, disrupting the feedback loop and stabilizing the shear layer. We investigated how quickly the suppression of fluctuations can be achieved.

The dominant frequency associated with the vortex convection is identified to be Rossiter Mode IV, obtained from the mid-cavity pressure sensor in the shear layer. We identified the phase space describing the dominant vortex convection using the phase and radius variables by projecting the spanwise averaged pressure fields onto the spatial mode associated with Rossiter Mode IV, obtained from dynamic mode decomposition. We obtained a trajectory with a center of rotation in the DMD-based phase space due to the flow's chaotic nature. The radius variable in the phase space is correlated with the strength of the spanwise vortices that formed in the shear layer, while the phase variable tracked the formation and the convection of the large-scale spanwise vortex due to the shear layer instability.

Using insights from resolvent analysis \cite{liu2021unsteady}, we considered spanwise actuation wavenumbers of $\beta=\pi$ and $2\pi$ for characterizing the phase-response and open-loop forcing as they are shown to be effective in suppressing pressure fluctuations when using harmonic actuation. Using the phase description, the phase sensitivity function corresponding to $\beta=\pi,2\pi$ is then computed by introducing a pulse train of impulses at the leading edge actuator and measuring the average shift in the frequency. The phase sensitivity function for the actuation wavenumbers revealed the most sensitive phase as the phase preceding the formation of the spanwise vortex at $x/D=2$, indicating that introducing perturbation at the leading edge at this phase would affect the formation of the large-scale spanwise structures.

Using the phase-sensitivity function, we analytically obtained the phase-sensitivity-based actuation waveform that can quickly modify the dominant frequency. Using these insights, we performed open-loop flow control simulations using the phase-sensitivity-based actuation waveform and a sinusoidal waveform at spanwise wavenumbers of $\beta=\pi$ and $2\pi$ and actuation frequencies of $\omega_f=0.8\omega_d,0.9\omega_d,1.1\omega_d$ and $1.2\omega_d$. All the considered controlled flows successfully suppressed the dominant tone at $St_L=1.32$, and we also observed that actuation using $\beta=2\pi$ introduces small structures that convect along the shear layer and promote the breakdown of the large-scale spanwise vortices in the cavity using both waveforms. For $\beta=\pi$, actuation using phase-sensitivity-based waveform achieved higher attenuation of pressure fluctuation when compared with sinusoidal actuation, especially at higher forcing frequencies. 

We also investigated how quickly the suppression of fluctuations could be achieved using both the waveforms for the three-dimensional actuation at $\beta=\pi$ and $2\pi$. We achieved a reduction of about 46\% by modifying the dominant frequency using three-dimensional actuation at $\beta=2\pi$ using sinusoidal waveform at $\omega=0.9\omega_d$ and the phase-sensitivity-based actuation waveform at $\omega=1.2\omega_d$ within about six convective time units. For $\beta=\pi$, the phase-sensitivity-based actuation waveform achieves a 10\% higher suppression than sinusoidal waveform with a maximum of 40\% reduction of pressure fluctuations within the cavity for $\omega_f=1.1\omega_d$. This reduction is also achieved within six convective time units from the start of actuation. At this wavenumber, the phase-sensitivity-based actuation waveform achieved fluctuation suppression by introducing streamwise vortices that promote earlier mixing and quickly result in the breakdown of the larger spanwise structures compared to the sinusoidal waveform, in accordance with the design of phase-based control. However, for $\beta=2\pi$, the sinusoidal actuation waveform achieved similar control performance to the phase-sensitivity-based actuation waveform and even outperformed the phase-sensitivity-based waveform for some scenarios due to the three-dimensional packing of the streamwise vortices introduced by the actuation. For $\beta=\pi$ and $2\pi$ phase-sensitivity-based actuation waveform achieves better reduction for actuation frequencies higher than the dominant frequency. The present findings indicate that the phase-based analysis can serve as a guideline to perform unsteady flow control through frequency modification of complex fluid flows. 
	
\section*{Acknowledgments}
	\label{sec:acknowledgments}
 V.G., L.S.U., L.N.C., and K.T. acknowledge the support from the US Air Force Office of Scientific Research (Grant numbers: FA9550-22-1-0013 and FA9550-21-1-0178) and the US National Science Foundation (Grant number: 2129639). Y.K. acknowledges the financial support from JSPS (Japan) KAKENHI (Grant numbers: JP24K06910 and JP20K03797). V.G. and K.T. thank Prof. Qiong Liu at New Mexico State University for sharing the details of the computational setup and resolvent-based control of cavity flows. V.G. also thanks Youngjae Kim and Dr. Kai Fukami at the University of California Los Angeles for helpful discussions on phase-based analysis.

	\section*{Declaration of interest}
	\label{sec:doi}
	The authors report no conflict of interest.

\appendix
\section*{Choice of phase space variables}
{\label{sec:appendix}}
We compare the phase and radius variables in the phase space obtained using $a(t)$ (equation~\ref{eq:DMD_coeff}) and $\dot{a}(t)$ with those obtained using Hilbert transform, $a-a^H$ (equation~\ref{eq:phase_hilbert}). Similar to the phase space computed using Hilbert transform shown in equation~\ref{eq:phase_hilbert}, we determine the phase space using normalized $a(t)$ and $\dot{a}(t)$ as 
\begin{equation}
    \theta_{\tan}(t)\equiv \tan^{-1}\left(\frac{a(t)}{\dot{a}(t)}\right)-\frac{\pi}{2},\quad r_{\tan}(t)\equiv\sqrt{\dot{a}(t)^2+a(t)^2}.
\end{equation}
Here $-\pi/2$ is added to $\theta_{\tan}(t)$ to account for the phase shift between the phase spaces of $a-a^H$ and $a-\dot{a}$. 

\begin{figure}
    \centering
    \includegraphics[width=0.95\textwidth]{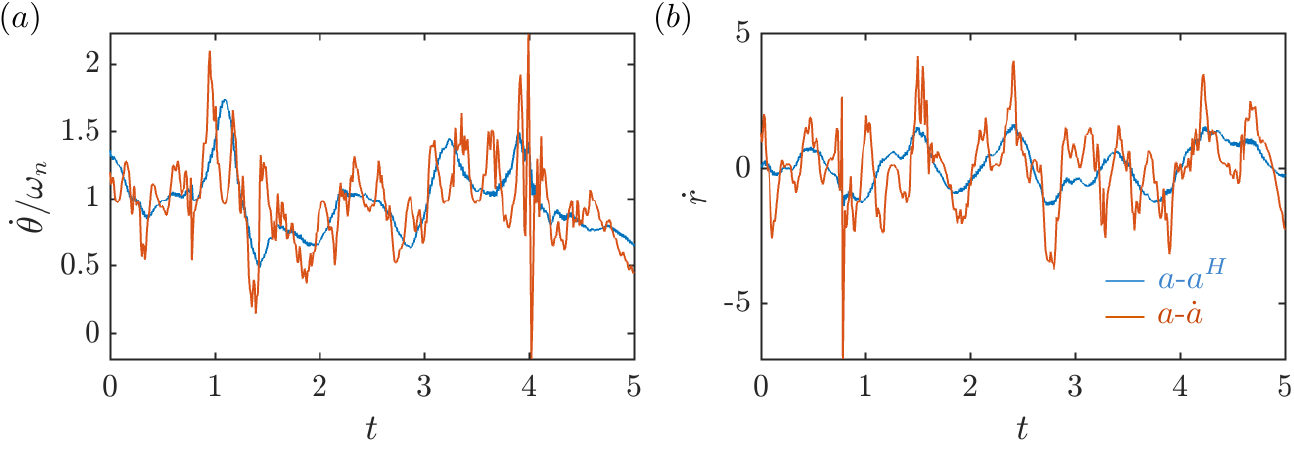}
    \caption{Phase dynamics $\dot{\theta}(t)$ and radial dynamics $\dot{r}(t)$ of phase space computed using Hilbert transform $a-a^H$, and time-derivative $a-\dot{a}$.}
    \label{fig:Hilbert_adot}
\end{figure}

The normalized phase dynamics $\dot{\theta}/\omega_n$ and radial dynamics $\dot{r}$ obtained using both the phase spaces $a-a^H$ and $a-\dot{a}$ are shown in figure~\ref{fig:Hilbert_adot}. We observe high-frequency fluctuations when the phase space is obtained using $a-\dot{a}$. On the other hand, the phase space evaluated using Hilbert transforms $a-a^H$ shows a smoother variation, filtering the high-frequency fluctuations. Hence, for the current study, we considered the Hilbert transform to define the phase and radial variables. In other systems that comprise lower fluctuation levels or those where the computation of Hilbert transform is not feasible, $a-\dot{a}$ phase space might be used due to its lower computational cost.

 \bibliographystyle{unsrtnat}
 \bibliography{refs}

\end{document}